\documentclass[usenatbib]{mn2e}

\usepackage{graphicx}
\usepackage{graphics}
\usepackage{amsmath}
\usepackage{amssymb}
\usepackage{color}
\usepackage{url}
\usepackage{hyperref}
\usepackage{pdflscape}
\usepackage{setspace}
\RequirePackage{fixltx2e}

\newcommand{\ion}[2]{#1$\;${\scshape{#2}}}

\def\aap{A\&A}

\def\apj{ApJ}
\def\apjl{ApJL}
\def\apjs{ApJS}

\def\mnras{MNRAS}

\def\araa{ARAA}
\def\pasp{PASP}
\def\apss{Ap\&SS}

\begin{document}

\title[High Signal-to-Noise Spectrum Toward J1009+0713]{A High Signal-to-Noise HST Spectrum Toward J1009+0713: Precise Absorption Measurements in the CGM of Two Galaxies}
\author[Lochhaas et al.]{Cassandra Lochhaas$^{1}$, Smita Mathur$^{1,2}$, Stephan Frank$^{1}$, Debopam Som$^{1}$,
\newauthor Yair Krongold$^{3}$, Varsha Kulkarni$^{4}$, David H. Weinberg$^{1,2}$, Fabrizio Nicastro$^{5,6}$, 
\newauthor Anjali Gupta$^{1}$ \\
$^{1}$ Department of Astronomy, The Ohio State University, 140 West 18th Avenue, Columbus, OH 43210, USA\\
$^{2}$ Center for Cosmology and Astroparticle Physics, The Ohio State University, 191 West Woodruff Avenue, Columbus, OH 43210, USA\\
$^{3}$ Instituto de Astronomia, Universidad Nacional Autonoma de Mexico, Apartado Postal 70264, 04510 CDMX, Mexico\\
$^{4}$ Department of Physics and Astronomy, University of South Carolina, Columbia, SC 29208, USA\\
$^{5}$ Osservatorio Astronomico di Roma --- INAF, Via di Frascati 33, I-00040 Monte Porzio Catone, RM, Italy\\
$^{6}$ Harvard-Smithsonian Center for Astrophysics, 60 Garden Street, Cambridge, MA 02138, USA}

\maketitle

\begin{abstract}
High signal-to-noise spectra toward background quasars are crucial for uncovering weak absorption in the circumgalactic medium (CGM) of intervening galaxies, such as the diagnostic lines of \ion{N}{v} that provide insight to the ionization process of warm gas but typically have low equivalent widths. We present a new spectrum from the \emph{Hubble} Space Telescope with a signal-to-noise ratio of $\sim20-35$ toward the quasar SDSS J1009+0713 and analyze absorption systems in the CGM of two $L^\star$ galaxies close to the line of sight. We identify additional absorption in the CGM of these galaxies that was not reported by the previous lower signal-to-noise spectrum, as well as Milky Way absorbers and quasar outflows from J1009+0713. We measure $\log (N_\mathrm{NV}/N_\mathrm{OVI})\sim-1.1$ for two CGM absorbers, inconsistent with gas in collisional ionization equilibrium and consistent with a radiatively cooling bulk flow of $\sim50-150$ km s$^{-1}$, which could be produced by galactic winds. These column density ratios are also consistent with those found for other $L^\star$ galaxies and for some gas in the Milky Way's halo. We place upper limits of $\log (N_\mathrm{NV}/N_\mathrm{OVI})<-1.8$ to $-1.2$ for other \ion{O}{vi} absorbers in the same halos, which suggests that \ion{O}{vi} is produced by different processes in different parts of the CGM, even within the same galactic halo. Together with the kinematically different structure of high- and low-ionization lines, these results indicate there are many components to a single galaxy's gaseous halo. We find the redshift number density of Ly-$\alpha$ forest absorbers and broad Ly-$\alpha$ absorbers are consistent with expectations at this redshift.
\end{abstract}


\section{Introduction}

The circumgalactic medium (CGM), the expanse of gas that fills a galaxy's halo outside of the galactic disk and inside the virial radius of the dark matter halo, is crucial for regulating gas flows into and out of the galaxy and as such is directly linked to galactic evolution. Measuring the density, temperature, total mass, and metallicity of gas in a galaxy's CGM is important for classifying it as inflowing or outflowing (although CGM gas need not be participating in bulk flows) and for predicting its origin, fate, and relation to galaxy properties \citep[for a review, see][]{Tumlinson2017}.

The extended CGM, far from the galaxy, is diffuse and difficult to measure in emission in the ultraviolet (UV) at low redshift and instead, there are many UV CGM surveys that focus on detecting CGM gas in absorption. In the X-ray band, the CGM is detected in both absorption and emission \citep[e.g.,][and references therein]{Gupta2012,Gupta2014}. ``Down the barrel" studies \citep{Heckman2000,Steidel2010,Bordoloi2011,Martin2012,Rubin2014,Chisholm2018} measure absorption against the starlight of the galaxy itself, but it can be difficult to separate absorption features due to the CGM from those due to the galaxy's interstellar medium. Quasar \\ \\ absorption line studies \citep[e.g.,][]{Chen1998,Kulkarni2005,Stocke2006,Meiring2009,Rudie2012,Tumlinson2013,Bordoloi2014,Kulkarni2015,Som2015,Keeney2017} use spectra of quasars that lie behind a galaxy of interest at some impact parameter to analyze absorption of the foreground galaxy's CGM gas in the quasar spectrum. The Cosmic Origins Spectrograph (COS) on the \emph{Hubble} Space Telescope has enabled large surveys of CGM gas surrounding several different types of galaxies, such as low-redshift $L^\star$ galaxies (COS-Halos; \citeauthor{Tumlinson2011} \citeyear{Tumlinson2011} and COS-GASS; \citeauthor{Borthakur2015} \citeyear{Borthakur2015}), starbursting galaxies \citep[COS-Burst][]{Heckman2017}, galaxies with active galactic nuclei \citep[COS-AGN][]{Berg2018}, luminous red galaxies \citep[COS-LRG][]{Chen2018}, and dwarf galaxies \citep[COS-Dwarfs][]{Bordoloi2014}. These surveys provide statistical trends of the CGM surrounding many types of galaxies, as functions of galaxy mass, halo mass, and impact parameter.

While useful for deducing typical properties of the CGM, many COS surveys are shallow and broad, focusing on obtaining as many spectra as possible with low signal-to-noise ratios (SNRs). Low, intermediate, and high ionization state metal lines are observed in the CGM \citep[e.g.,][]{Stocke2006,Wakker2009,Werk2013,Keeney2017}, which is usually interpreted as the CGM being home to a variety of gas temperatures, densities, and ionization processes \citep[but see, e.g.,][]{Stern2016,Stern2018}. High SNR spectra probing the CGM are necessary to obtain precise measurements of column densities and absorption line kinematics that promote detailed ionization modeling of the CGM gas. In addition to the increase in precision granted by higher SNR, deeper spectra may also identify weak lines that shallow spectra cannot: \ion{N}{v}, an ion that provides insights to the ionization process of warm gas, and broad Ly-$\alpha$ absorbers (BLAs), which indicate \ion{H}{i} gas at a high temperature $\sim10^5-10^6$ K \citep{Richter2006,Tepper-Garcia2012}. Accurately tracing the kinematics of CGM absorbers would be best done with high-resolution spectra, but in its absence, high-SNR spectra can provide some increase in precision that low-SNR lacks.

In particular, the high-ionization state metal lines \ion{N}{v} and \ion{O}{vi} could arise due to a number of processes: gas in collisional ionization equilibrium (CIE), gas in photoionization equilibrium, gas in non-equilibrium cooling flows, turbulent mixing, gas near active galactic nuclei \citep{Segers2017}, or hot gas shocking cold clouds \citep{Indebetouw2004,Wakker2012,Bordoloi2017}. Recent papers argue that \ion{O}{vi}-tracing gas arises in radiatively cooling flows of $T\sim10^{5.5}$ K gas \citep{Werk2016,McQuinn2018} or in a $T\sim10^4$ K photoionized volume-filling gaseous reservoir within halos \citep{Stern2016,Stern2018}. Precise measurements of the \ion{O}{vi} and \ion{N}{v} column densities will help discriminate between models to determine where the high-ionization state metals originate and what phase of gas they trace \citep{Werk2016}. In particular, \ion{N}{v} typically has lower column densities than \ion{O}{vi} by a factor of ten \citep{Werk2016}, so a high SNR spectrum can be necessary to detect it.

In this paper, we present a spectrum of the quasar SDSS J100902.06+071343.9 (hereafter abbreviated as J1009+0713) with a SNR of $20-35$ per resolution element. This quasar, at a redshift of $z=0.456$, illuminates the CGM of two foreground galaxies at lower redshifts. We describe the reduction and normalization of the spectrum and present the full spectrum with all identified absorption lines in \S\ref{sec:spectrum}. We fit all detected absorption lines at the two galaxy redshifts and report column densities, velocity offsets of absorption from each galaxy's systemic velocity, and Doppler broadening parameters in \S\ref{sec:fitting}, and also discuss improvements to measurements over the previous spectrum of this object in the COS-Halos survey. Section~\ref{sec:discussion} presents an estimate of the ionization processes most likely to produce the low- and high-ionization absorption (\S\ref{sec:ionization}) and specifically the detected \ion{N}{v} and \ion{O}{vi} absorption (\S\ref{sec:NV_OVI}). We summarize and conclude in \S\ref{sec:summary}. We find that the number of Ly-$\alpha$ absorbers in the spectrum matches the expected number of Ly-$\alpha$ absorbers for the covered redshift range in Appendix \ref{sec:Lya_forest} and identify which of those are broad Ly-$\alpha$ absorbers in Appendix \ref{sec:BLA}.

\section{The Spectrum}
\label{sec:spectrum}

\begin{figure*}
\begin{minipage}{175mm}
\centering
\includegraphics[width=\linewidth]{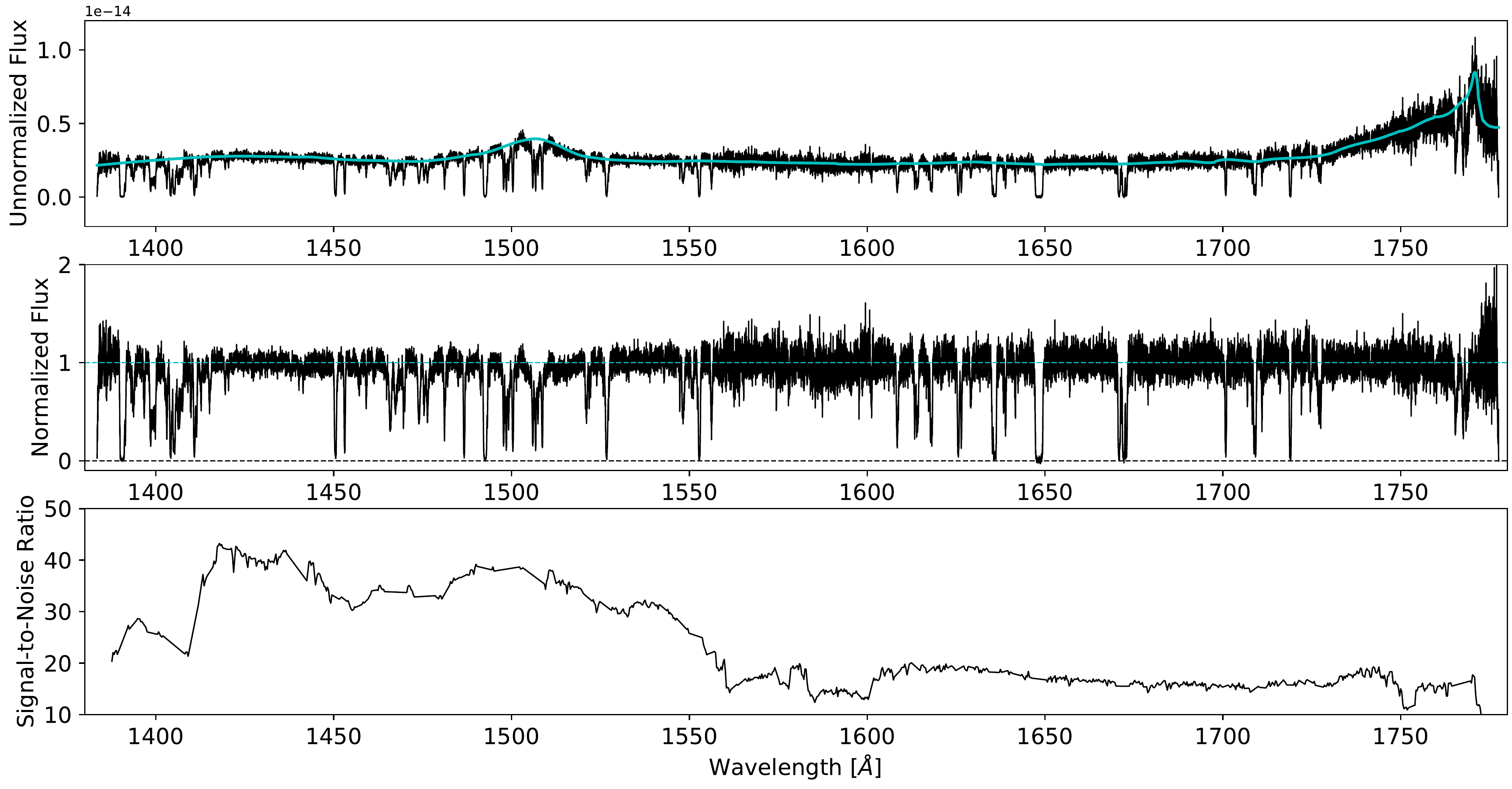}
\caption{Top: The full, unnormalized spectrum. The cyan curve shows the best-fit continuum curve. Middle: The normalized spectrum, after dividing by the cyan continuum curve in the top panel. Bottom: The SNR per resolution element across the full spectrum.}
\label{fig:normalization}
\end{minipage}
\end{figure*}

The quasar J1009+0713 was chosen for this study because it is bright, with a GALEX FUV magnitude of $18.08\pm0.09$, making a high SNR spectrum easier to obtain, and because the two galaxies' CGM it probes were previously measured to have high \ion{O}{vi} column densities, indicating the presence of substantial warm-hot gas. The previously obtained data are also useful for determining the degree of improvement a deeper spectrum can provide. The \emph{Hubble} Space Telescope observed J1009+0713 for 21 orbits in Cycle 24, a total exposure time of 69569 seconds ($\approx$1159 minutes), using the Cosmic Origins Spectrograph (COS) 160M grating. Half of the exposure time used the 1577 \AA\ central wavelength of the grating and the other half used the 1600 \AA\ central wavelength, to cover the gap in detector segments. Near the absorption lines of interest, we find a minimum SNR per resolution element of $\sim15$ and a typical SNR per resolution element of $20-35$ (see Figure~\ref{fig:normalization}), much higher than the previously obtained spectrum of this object in the COS-Halos survey \citep{Tumlinson2013}, which had a SNR of $7-15$ per resolution element. Converting the signal to noise ratio per resolution element to equivalent width (EW) limits produces $3\sigma$ EW limits ranging from $0.006$ to $0.029$ \AA\ over the spectrum, with an average EW $3\sigma$ limit of $0.015$ \AA, much better than the COS-Halos spectrum $2\sigma$ EW limit of $0.06$ \AA\ \citep{Werk2013}. On the wavelength range investigated, the spectral resolution, measured as the FWHM of the line spread function, is $\sim0.1$ \AA. The data were reduced and combined into a single spectrum using the standard CALCOS pipeline. The continuum flux of the target in the range of the spectrum from $\sim1380$ \AA\ to $\sim1775$ \AA\ is $\sim3.5\times10^{-15}$ ergs s$^{-1}$ cm$^{-2}$ \AA$^{-1}$, and the typical error on the flux is $\sim3\times10^{-16}$ ergs s$^{-1}$ cm$^{-2}$ \AA$^{-1}$ at the continuum level, dropping to $1\times10^{-16}$ ergs s$^{-1}$ cm$^{-2}$ \AA$^{-1}$ in absorption features.


We normalize the spectrum using an automatic, iterative process. In essence, we first smooth the spectrum over a large number of pixels (on the order of 600). Then, we subdivide the spectrum into small bins of $30-40$ pixels, the midpoints of which serve as fulcrums for a spline-fitting of a third order polynomial. In the first iteration, the median of all pixels within each bin is taken as fiducial value for the estimate of the continuum. We then reject in each bin pixels that lie below the spline fit by more than $2\sigma$, thereby eliminating most areas that are affected by absorption, and subsequently recalculate the median for the remaining pixels, thus producing a new estimate. This process is iterated until the change in the resulting continuum estimate falls below the $3\%$ margin for all remaining pixels. For more details of this procedure see e.g. \citet{Pieri2014} and \citet{Frank2018}, both of which have employed the same method. Figure~\ref{fig:normalization} shows the unnormalized spectrum in the top panel and the normalized spectrum in the middle panel. Also plotted in the top panel is the continuum estimation as a cyan curve.

We calculate the SNR per resolution element by first masking absorption in the normalized spectrum by-eye. Using the remaining (continuum-only) pixels, we bin the normalized spectrum into 25-pixel bins and calculate the ratio of the flux to the flux error in each pixel within the bin, then find the mean of this ratio across all pixels in the bin. The resultant mean value in each bin is then multiplied by $\sqrt{8}$ to obtain the SNR per resolution element, as COS has roughly 8 pixels per resolution element. The SNR per resolution element is highly wavelength-dependent and is plotted in the bottom panel of Figure~\ref{fig:normalization}. We then calculate the limiting EW within the 25-pixel bins as the central wavelength of the bin over the mean resolution of the spectrum, $R=16500$, divided by the SNR in the bin.

\subsection{Identifying Absorption Lines}
\label{sec:lines}


The full, normalized spectrum is plotted in 25-\AA\ segments in Figures~\ref{fig:spectrum_chunks1}-\ref{fig:spectrum_chunks5}. We identify the wavelength of potential absorption lines using the atomic data provided with the VPFit\footnote{R. F. Carswell \& J. K. Webb, https://www.ast.cam.ac.uk/\~rfc/vpfit.html} program, most of which was compiled in \citet{Morton2003}, at the redshift of the two galaxies of interest \citep[as determined by][]{Werk2012}, as well as the redshift of the quasar ($z=0.456$), the Milky Way ($z=0$), and a damped Lyman-$\alpha$ absorber (DLA) in this line of sight identified by \citet[][$z=0.1140$]{Meiring2011}. There are many absorption lines that are not identified in this way, which we assume to be hydrogen absorption due to the Ly-$\alpha$ forest (see \S\ref{sec:Lya_forest}) at low redshift or the broad Ly-$\alpha$ absorption lines.

In addition to identifying absorption at or near atomic transitions, we search for absorption blue-shifted from the quasar's redshift, indicative of quasar outflows. We identified the repeating, multi-component structure at $\sim 1497-1502$ \AA\ and $\sim 1505-1510$ \AA\ as highly-blueshifted \ion{O}{vi} $\lambda1031$ and \ion{O}{vi} $\lambda1037$ absorption from the quasar's outflow, which corresponds to velocities of $\sim100-1300$ km s$^{-1}$. Using the extent of the \ion{O}{vi} $\lambda1031$ absorption as indicative of the velocities over which we could expect to see absorption from the quasar's outflow, we searched for other absorption systems at these velocity blue-shifts from the quasar's redshift. Other quasar outflow absorption lines we find are \ion{C}{iii} $\lambda977$ at $\sim1419$ \AA\ and $\sim1421$ \AA. The regions of quasar outflow absorption are marked in Figures~\ref{fig:spectrum_chunks1}-\ref{fig:spectrum_chunks5} as horizontal cyan lines extending over the full range of the \ion{O}{vi} $\lambda1031$ absorption, and labeled as ``outflow."

\begin{figure*}
\begin{minipage}{175mm}
\centering
\includegraphics[width=\linewidth]{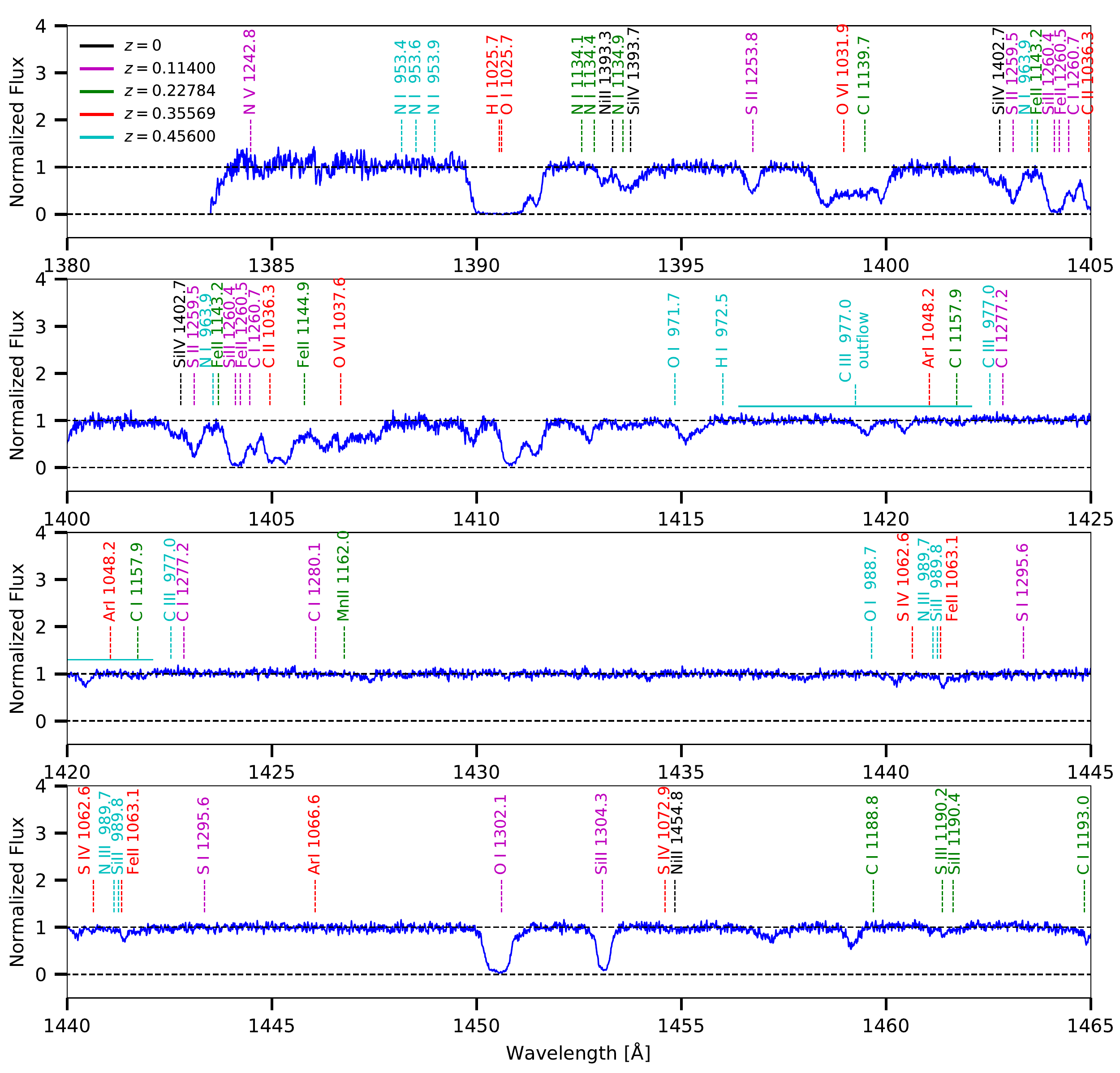}
\caption{A segment of the normalized spectrum with the location of common absorption lines identified at the redshifts of the Milky Way ($z=0$, black), the quasar ($z=0.456$, cyan), galaxy 1 ($z=0.22784$, green), galaxy 2 ($z=0.35569$, red), and the DLA identified by \citet{Meiring2011} ($z=0.1140$, magenta). Transitions with oscillator strengths $>0.01$ of elements H, He, C, N, O, Na, Mg, Si, S, Fe, Ar, P, Mn, and Ni are shown. Regions of the spectrum where we might expect to find highly-blueshifted absorption from the quasar outflow are marked by horizontal cyan lines that cover the full velocity range of the quasar outflow's \ion{O}{vi} $\lambda1031$ absorption. There are several unidentified lines in the spectrum, which are reported as such in Table~\ref{tab:unidentified_lines}; most of these are likely to be intervening Ly-$\alpha$ or Ly-$\beta$ lines. Rows show consecutive 25 \AA\ segments, including the last 5 \AA\ of the previous segment to allow for close examination of features near the edges of each panel.}
\label{fig:spectrum_chunks1}
\end{minipage}
\end{figure*}

\begin{figure*}
\begin{minipage}{175mm}
\centering
\includegraphics[width=\linewidth]{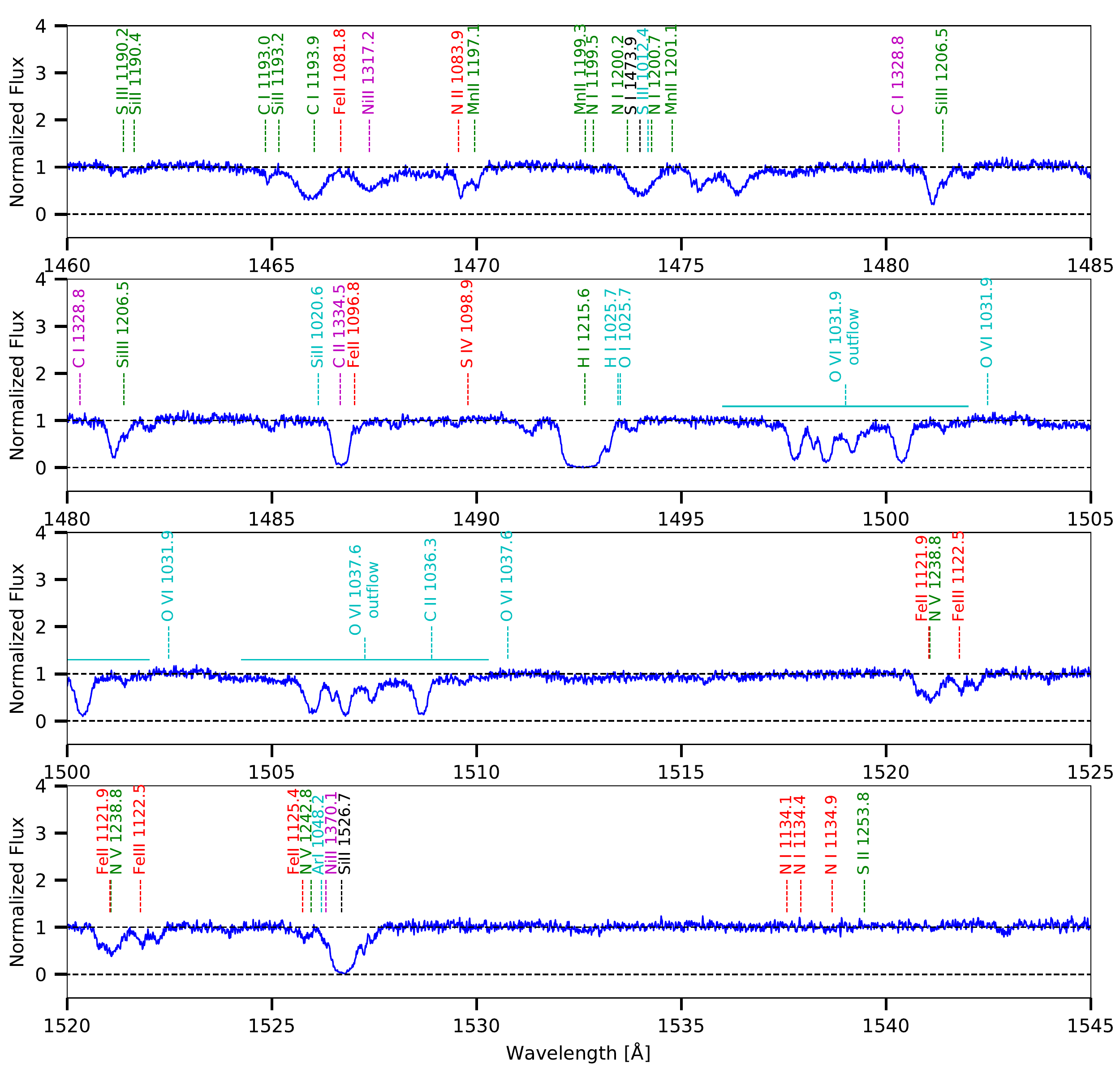}
\caption{Continuation of the spectrum shown in Figure~\ref{fig:spectrum_chunks1}.}
\label{fig:spectrum_chunks2}
\end{minipage}
\end{figure*}

\begin{figure*}
\begin{minipage}{175mm}
\centering
\includegraphics[width=\linewidth]{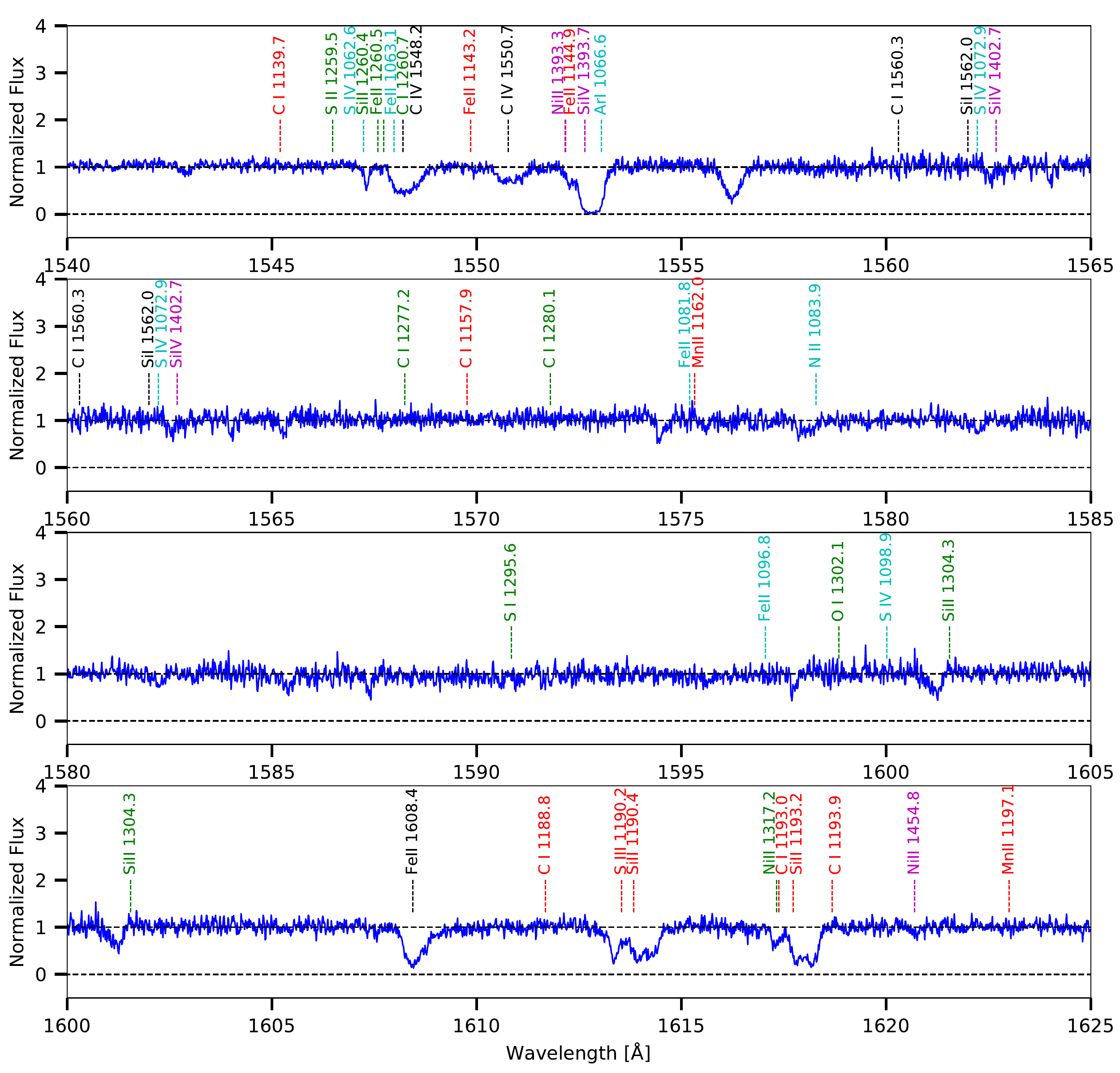}
\caption{Continuation of the spectrum shown in Figures~\ref{fig:spectrum_chunks1}-\ref{fig:spectrum_chunks2}.}
\label{fig:spectrum_chunks3}
\end{minipage}
\end{figure*}

\begin{figure*}
\begin{minipage}{175mm}
\centering
\includegraphics[width=\linewidth]{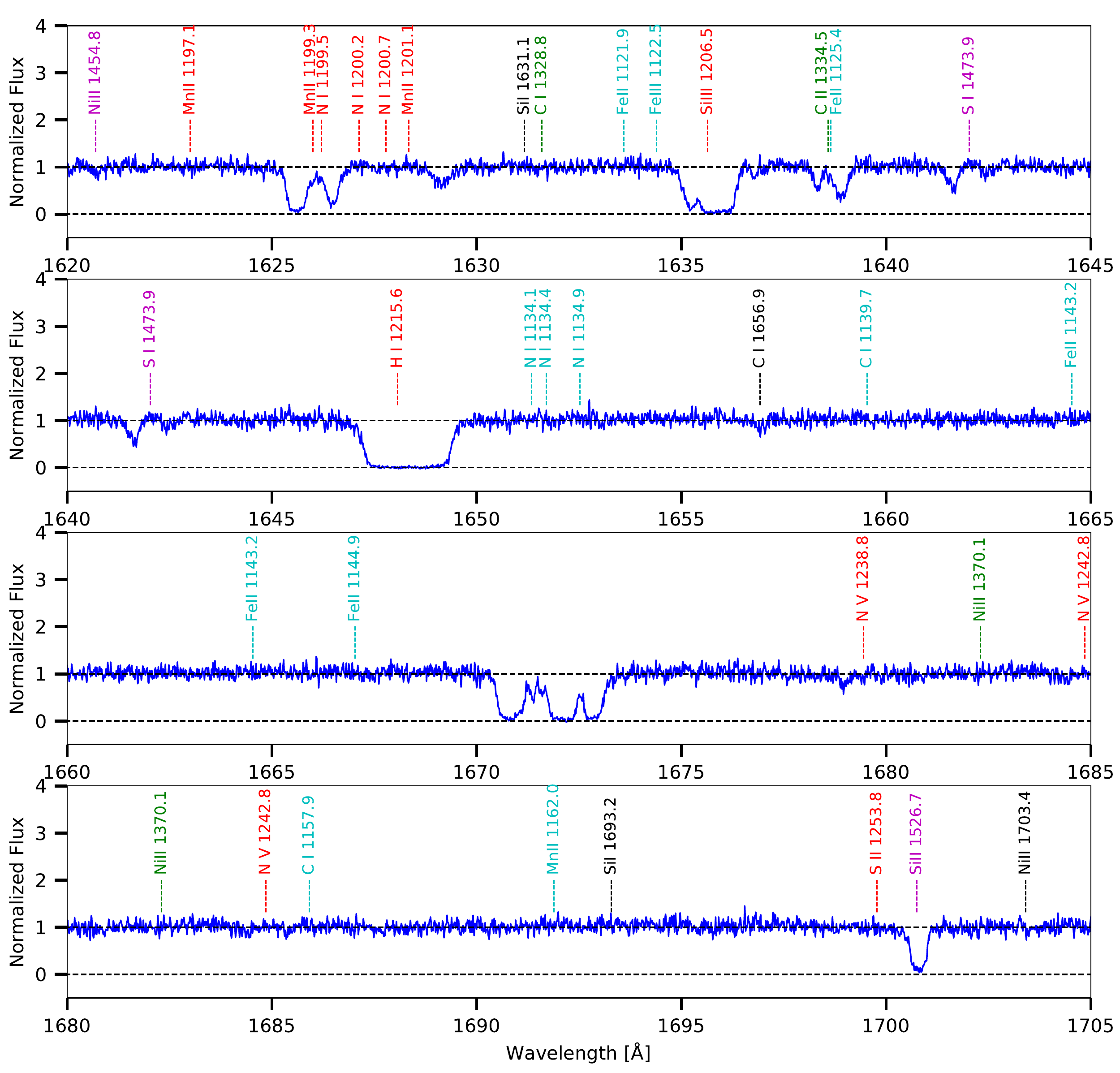}
\caption{Continuation of the spectrum shown in Figures~\ref{fig:spectrum_chunks1}-\ref{fig:spectrum_chunks3}.}
\label{fig:spectrum_chunks4}
\end{minipage}
\end{figure*}

\begin{figure*}
\begin{minipage}{175mm}
\centering
\includegraphics[width=\linewidth]{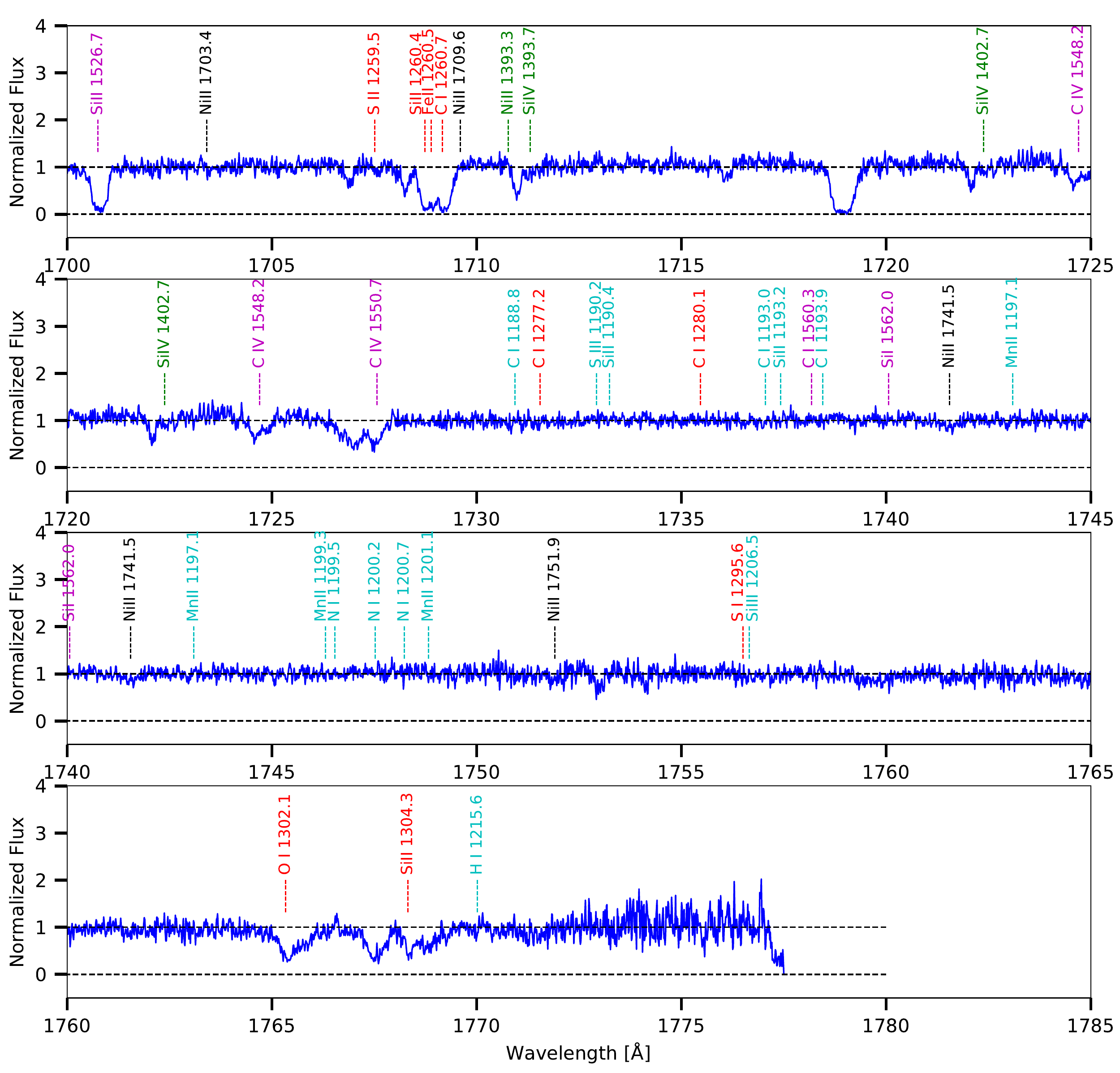}
\caption{Continuation of the spectrum shown in Figures~\ref{fig:spectrum_chunks1}-\ref{fig:spectrum_chunks4}.}
\label{fig:spectrum_chunks5}
\end{minipage}
\end{figure*}

Table~\ref{tab:identified_lines} lists the equivalent widths of all lines we identify as belonging to the Milky Way (MW), the quasar (QSO), the DLA identified by \citet{Meiring2011}, or the combination of several lines from the MW, QSO, DLA, or the CGM of galaxies 1 or 2 that may be too blended to disentangle and fit. We also include lines that are highly-blueshifted absorption from the quasar outflow and the range of velocities over which absorption occurs. Table~\ref{tab:identified_lines} lists only those lines that we identify but do not fit, Table~\ref{tab:unidentified_lines} lists all absorbing systems that we do not identify, and Tables~\ref{tab:fits1} and~\ref{tab:fits2} list lines that we identify as arising from the CGM of the two galaxies and which we fit. We discuss the unidentified absorption lines presented in Table~\ref{tab:unidentified_lines} and calculate their d$N$/d$z$ under the assumption that they are primarily Ly-$\alpha$ forest lines (in Appendix~\ref{sec:Lya_forest}) or BLAs (in Appendix~\ref{sec:BLA}).

We consider an identified absorption system to be blended or contaminated by another when the absorption component structure of one transition of a particular ion does not match that of other transitions of the same ion. In all such cases, we simultaneously fit all transitions of the ion but include additional absorption components for intervening lines in the fit. Of course, ions for which our spectrum contains just one transition may also be blended with intervening absorbers, but in this case we simply assume that all detected absorption is due to the transition of interest. Note that this assumption may mis-identify the handful of CGM absorbing ions for which we measure only a single transition, but we adopt this assumption regardless for easier comparison with the COS-Halos study, which also used this assumption. For certain contaminated lines, it is clear that another absorption line nearby is responsible for blending, so we include identified nearby lines in Table~\ref{tab:identified_lines}. In the cases where we cannot identify the line responsible for blending, we assume the blending may be due to absorption in the Ly-$\alpha$ forest, and list these unidentified absorption features in Table~\ref{tab:unidentified_lines}.

Lines that we identify as belonging to the CGM of either galaxy 1 or galaxy 2 are fit and investigated further in Section~\ref{sec:fitting}. Sections~\ref{sec:gal1_overview} and~\ref{sec:gal2_overview} first give a brief overview of these galaxies and the lines we identify in their CGM.

\begin{table}
\centering
\begin{tabular}{l c c}
 &  & Equivalent \\
Ion$^\mathrm{a}$ & System$^\mathrm{b}$ & width$^\mathrm{c}$ (\AA)\\
\hline
*\ion{Si}{iv} 1393 & MW & 0.40 \\
~~+\ion{N}{i} 1134 & ~~+gal1 & \\
\ion{S}{ii} 1253 & DLA & 0.17 \\
*\ion{Si}{iv} 1402$^\dag$ & MW & 0.36 \\
~~+\ion{S}{ii} 1259 & ~~+DLA & \\
*\ion{Si}{ii} 1260$^{\dag \ddag}$ & DLA & 1.72\\
~~+\ion{C}{ii} 1036 & ~~+gal2 & \\
~~+\ion{Fe}{ii} 1144 & ~~+gal1 & \\
~~+\ion{O}{vi} 1037 & ~~+gal2 & \\
\ion{H}{i} 972$^\dag$ & QSO & 0.20 \\
\ion{C}{iii} 977 & QSO outflow at $\sim-600$ km/s & 0.11 \\
\ion{C}{iii} 977 & QSO outflow at $\sim-440$ km/s & 0.06 \\
\ion{S}{iv} 1062$^\dag$ & gal2 & 0.06 \\
\ion{Fe}{ii} 1063$^\dag$ & gal2 & 0.08 \\
\ion{O}{i} 1302 & DLA & 0.71 \\
\ion{Si}{ii} 1304 & DLA & 0.36 \\
*\ion{C}{i} 1193$^\dag$ & gal1 & 0.55 \\
~~+\ion{Fe}{ii} 1081 & ~~+gal2 & \\
\ion{Ni}{ii} 1317$^{\dag \ddag}$ & DLA & 0.41 \\
*\ion{S}{i} 1473$^\dag$ & MW & 0.43 \\
~~+\ion{S}{iii} 1012 & ~~+QSO & \\
*\ion{C}{ii} 1334 & DLA & 0.50 \\
~~+\ion{Fe}{ii} 1096 & ~~+gal2 & \\
\ion{O}{vi} 1031 & QSO outflow & 1.40 \\
 & from $\sim-1000$ to $\sim-200$ km/s & \\
\ion{O}{vi} 1037 & QSO outflow & 1.69 \\
 & from $\sim-1000$ to $\sim-200$ km/s & \\
*\ion{Ni}{ii} 1370$^\dag$ & DLA & 0.80 \\
~~+\ion{Si}{ii} 1526 & ~~+MW & \\
\ion{C}{iv} 1548$^\dag$ & MW & 0.39 \\
\ion{C}{iv} 1550 & MW & 0.20 \\
\ion{Si}{iv} 1393$^\dag$ & DLA & 0.70 \\
\ion{Si}{iv} 1402$^\dag$ & DLA & 0.06 \\
\ion{N}{ii} 1083 & QSO & 0.13 \\
\ion{Fe}{ii} 1608 & MW & 0.52 \\
\ion{C}{i} 1656 & MW & 0.04 \\
\ion{Si}{ii} 1526 & DLA & 0.43 \\
\ion{S}{ii} 1259 & gal2 & 0.04 \\
\ion{C}{iv} 1548 & DLA & 0.12 \\
\ion{C}{iv} 1550$^{\dag\ddag}$ & DLA & 0.09
\end{tabular}
\caption{$^\mathrm{a}$Identified absorption lines listed by ion and wavelength. $^\mathrm{b}$The system the absorber belongs to: the Milky Way (MW), the quasar (QSO), the DLA at $z=0.1140$, or galaxy 1 or 2 (gal1 or gal2). $^\mathrm{c}$The equivalent width of the full absorption profile that includes this transition. $^*$These absorbers contain blended transitions belonging to multiple systems, so we report the equivalent width of the full profile and list the other lines and systems in the blend with `+' in the rows immediately below. $^\dag$These absorbers are located near the wavelength of multiple transitions so it is not clear which transition is responsible for absorption. We list them as the nearest transition with the strongest oscillator strength, but caution our identification may be incorrect. $^\ddag$These absorbers are blended with other transitions of interest so we fit the full multicomponent absorption system and then calculate the equivalent width of the fit for the component listed, instead of calculating the equivalent width from the data (see text).}
\label{tab:identified_lines}
\end{table}

\begin{table}
\centering
\begin{tabular}{l c c c}
Central & Equivalent & & Column\\
wavelength$^\mathrm{a}$ (\AA) & width$^\mathrm{b}$ (\AA) & Redshift$^\mathrm{c}$ & density$^\mathrm{d}$ (log cm$^{-2}$)\\
\hline
1409.9$^\ddag$ & 0.12 &  & \\
1411.1$^\ddag$ & 0.75 &  & \\
1412.7 & 0.11 & 0.16209 & 13.32\\
1457.2$^\ddag$ & 0.15 &  & \\
1459.2* & 0.13 & 0.20029 & 13.40\\
{\bf 1466.0*$^\dag$} & {\bf 0.54} & {\bf 0.20585} & {\bf 14.07}\\
{\bf 1468.6*$^\dag$} & {\bf 0.26} & {\bf 0.20821} & {\bf 13.63}\\
{\bf 1475.5} & {\bf 0.25} & {\bf 0.21371} & {\bf 13.68}\\
{\bf 1476.4} & {\bf 0.34} & {\bf 0.21444} & {\bf 13.80}\\
{\bf 1477.6} & {\bf 0.10} & {\bf 0.21544} & {\bf 13.31}\\
1482.0*$^\dag$ & 0.05 & 0.21907 & 12.94\\
1485.0 & 0.05 & 0.22151 & 12.99\\
1491.3*$^{\dag\ddag}$ & 0.11 &  & \\
1493.8*$^\dag$ & 0.06 & 0.22877 & 13.02\\
{\bf 1521.5*$^\dag$} & 0.42 & 0.25095 & 12.95\\
 & & {\bf 0.25129} & {\bf 13.56}\\
 & & 0.25183 & 13.29\\
 & & 0.25214 & 13.13\\
{\bf 1556.3} & {\bf 0.30} & {\bf 0.28014} & {\bf 13.84}\\
{\bf 1613.3*$^\dag$} & 0.46 & 0.32712 & 13.50\\
 & & {\bf 0.32745} & {\bf 13.76}\\
1626.0* & 0.84 & 0.33721 & 14.35\\
 & & 0.33791 & 13.88\\
{\bf 1629.1} & {\bf 0.15} & {\bf 0.34011} & {\bf 13.41}\\
1636.8*$^\dag$ & 0.03 & 0.34639 & 12.73\\
1638.9*$^\dag$ & 0.26 & 0.34812 & 13.72\\
1641.6* & 0.12 & 0.35037 & 13.39\\
1671.8 & 2.04 & 0.37440 & 14.78\\
 & & 0.37485 & 13.31\\
 & & 0.37502 & 12.99\\
 & & 0.37544 & 15.18\\
 & & 0.37606 & 14.40\\
1706.9 & 0.11 & 0.40406 & 13.27\\
1716.1 & 0.05 & 0.41163 & 13.01\\
1718.9 & 0.62 & 0.41398 & 14.63\\
{\bf 1727.0*$^\dag$} & 0.44 & {\bf 0.42057} & {\bf 13.70}\\
 & & 0.42105 & 13.32\\
1767.5*$^\dag$ & 0.34 & 0.45394 & 13.81\\
{\bf 1768.4*$^\dag$} & 0.35 & 0.45461 & 13.30\\
 & & {\bf 0.45493} & {\bf 13.61}\\
\end{tabular}
\caption{Unidentified absorbers in the spectrum. $^\mathrm{a}$The approximate central wavelength of the absorbing structure. $^\mathrm{b}$The equivalent width of the full absorbing structure. $^\mathrm{c}$The redshift of the centroid of each component within the absorber, assuming the absorber is a Ly-$\alpha$ transition. $^\mathrm{d}$The log of the best-fit column density for each absorption component, assuming the transition is Ly-$\alpha$. Lines that are later identified as broad Ly-$\alpha$ absorbers (BLAs) are marked in bold (see Appendix \ref{sec:BLA}). *These absorbers are located near the wavelengths of metal line transitions that we identify, so they may be red- or blueshifted metal lines instead of Ly-$\alpha$ absorbers. However, we report them here as unidentified absorbers either because they do not align perfectly with the wavelength of nearby metal lines or because they are much stronger than the oscillation strength of the nearby metal lines would imply the absorption to be. $^\dag$These absorbers are blended with other metal line transitions of interest that we fit, so we calculate the equivalent width using the best fit of the unidentified component instead of calculating it directly from the data (see text). $^\ddag$These absorbers are consistent with being Ly-$\beta$ absorption, rather than Ly-$\alpha$ like we assume for other unidentified absorbers (see Appendix~\ref{sec:Lya_forest}), so we do not report their redshift or column density.}
\label{tab:unidentified_lines}
\end{table}

\subsubsection{Galaxy 1 at $z=0.22784$}
\label{sec:gal1_overview}

The first galaxy is located 59 kpc from the quasar sightline at a redshift of $z=0.22784$, measured from the Keck LRIS optical spectra taken as part of the COS-Halos program \citep{Werk2012}. It has a stellar mass of $10^{9.85} M_\odot$ and a star formation rate of $4-9\ M_\odot$ yr$^{-1}$, depending on whether it was measured from the Balmer emission lines or [\ion{O}{ii}], with the [\ion{O}{ii}]-measured star formation rate being higher.

Common CGM absorption lines covered by our spectrum for the first galaxy are \ion{H}{i} $\lambda1215$ \AA, \ion{C}{ii} $\lambda1334$ \AA, \ion{N}{v} $\lambda1238,\lambda1242$ \AA, \ion{Si}{ii} $\lambda1190,\lambda1193,\lambda1260$ \AA, \ion{Si}{iii} $\lambda 1206$ \AA, and \ion{Si}{iv} $\lambda1393,\lambda1402$ \AA. Of these, COS-Halos had previously identified \ion{C}{ii} $\lambda1334$ \AA\ and \ion{Si}{iii} $\lambda1206$ \AA\ only. The COS-Halos spectrum also covers \ion{O}{vi} $\lambda1031,\lambda1037$ \AA\ whereas our spectrum does not, so we include these lines in our analysis by performing our own fits to the COS-Halos spectrum, for consistency. Note, however, that the COS-Halos spectrum has a much lower SNR than the new spectrum reported here, so any best-fit parameters we derive from it have significantly larger errors than the fits to absorption in the new spectrum.

\subsubsection{Galaxy 2 at $z=0.35569$}
\label{sec:gal2_overview}

Galaxy 2 is located 43 kpc from the quasar sightline, at a redshift of $z=0.35569$, and has a stellar mass of $10^{10.24}M_\odot$ and a star formation rate of $2-3 M_\odot$ yr$^{-1}$, measured from both the Balmer emission lines and [\ion{O}{ii}] emission \citep{Werk2012}.

The absorption lines we identify in the CGM of the second galaxy at $z=0.35569$ are \ion{H}{i} $\lambda1215,\lambda1025$ \AA, \ion{N}{ii} $\lambda1084$ \AA, \ion{N}{v} $\lambda 1238,\lambda1242$ \AA, \ion{O}{i} $\lambda1302$ \AA, \ion{O}{vi} $\lambda1031,\lambda1037$ \AA, \ion{Si}{ii} $\lambda 1190,\lambda1193,\lambda1260,\lambda1304$ \AA, and \ion{Si}{iii} $\lambda1206$ \AA. Of these, COS-Halos had previously identified \ion{N}{ii} $\lambda1084$ \AA, \ion{O}{i} $\lambda1302$ \AA, \ion{O}{vi} $\lambda1031$ \AA, \ion{Si}{ii} $\lambda1260$ \AA, and \ion{Si}{iii} $\lambda1206$ \AA.

\subsection{Newly-Identified Absorption}
\label{sec:new_lines}

Due to the high SNR of our spectrum, we have identified new absorption components that were not previously identified in the COS-Halos study of this spectrum. For both galaxies, the most important new discovery is the \ion{N}{v} absorption, as this high ion is a useful diagnostic of the ionization processes of the gas traced by high ions, especially when combined with \ion{O}{vi} (see \S\ref{sec:NV_OVI}).

The only metal absorption lines identified by both the previous COS-Halos spectrum and our spectrum for galaxy 1 are \ion{C}{ii} $\lambda1036$ \AA\ and \ion{Si}{iii} $\lambda1206$ \AA. We identify the same absorption components for these two metal ions as COS-Halos. We find an additional, weak absorption component redshifted from the \ion{Si}{iii} line that was not detected by COS-Halos, however we do not assume this absorption is due to \ion{Si}{iii} because this absorption structure is not observed in any of the other low-ion absorption. We list it as unidentified absorption in Table~\ref{tab:unidentified_lines}. The \ion{Si}{ii} lines and \ion{N}{v} $\lambda1242$ \AA, the one \ion{N}{v} transition that is not blended with intervening absorption (see \S\ref{sec:gal1_fits}), that we identify have equivalent widths of $0.03-0.07$ \AA\, which fall around or below the EW limit in the COS-Halos spectrum of $\sim0.06$ \AA\ and thus were not previously reported \citep{Werk2013}.

For the ions detected in absorption by COS-Halos in the CGM of galaxy 2, we do not detect any additional absorption components that they do not report. However, our higher SNR spectrum allows for a cleaner decomposition of absorption into components. For example, our detected absorption in \ion{O}{vi} is spread over a similar velocity range as determined by \citet{Tumlinson2011}, but the most blueshifted and most redshifted velocity absorption components are more clearly defined and fits produce smaller errors on velocity centroids. The same is true of our \ion{Si}{ii} and \ion{N}{ii} lines. We report a firm detection of weak \ion{N}{v} absorption that was not seen in the COS-Halos spectrum, due to being weaker than the COS-Halos spectrum detection limit of $\sim0.06$ \AA. We also report absorption in more transitions of \ion{Si}{ii} than COS-Halos reported, allowing for a better fit of this ion's absorption.

We now see one of the benefits of high SNR spectra like the one presented in this paper. By pushing to a deeper EW limit than previously explored, we can probe weak absorbers that would otherwise go undetected in broad, shallow surveys. Such weak absorbers can help classify the column density distribution of gas in the CGM. In the CGM of the two galaxies located along this line of sight, we detect previously-undetected low-EW absorption components: two in the CGM of galaxy 1 and one in the CGM of galaxy 2, despite a factor of $\sim6$ improvement in the EW limit of our spectrum over COS-Halos spectra. Note that while the average EW limit of our spectrum is a factor of $\sim4$ improvement over COS-Halos, the EW limit is wavelength-dependent and is a factor of $\sim6$ improvement at the wavelengths of the majority of the lines in both galaxies' CGM. Naively, we might expect to find a factor of 6 more previously undetected absorption components, instead of the $\sim30\%$ and $\sim6\%$ more that we find for galaxies 1 and 2, respectively. This implies either that these small-EW absorbers are impossible to disentangle because they have similar velocities as the large-EW absorbers, or that large-EW absorbers are made up of many small-EW absorbers, or that the number of absorbers simply does not continue increasing to smaller EWs. The former two arguments imply that most gas in the CGM has similar velocities despite being distributed among many ``clouds" with small column densities and the latter implies that most absorbing gas in the CGM is located in large clouds (or simply a volume-filling gaseous halo) with large column densities, or in small, but very dense clouds. In either case, the high-SNR spectrum was necessary to determine that low-SNR spectrum are not missing much CGM absorption. This suggests there may be a turnover in the column density distribution of absorbers at low EW that low SNR spectra cannot probe.

\section{Voigt Profile Fitting}
\label{sec:fitting}

We use the software VPFit to fit multicomponent Voigt profiles to the absorption lines identified in \S\ref{sec:lines} as belonging to the CGM of galaxies 1 and 2. The non-Gaussian line spread function (LSF) of COS is passed to VPFit and handled self-consistently, such that the LSF is considered and accounted-for in the errors on the fits reported by VPFit. We obtain velocity centroids relative to the galaxy systemic velocity $v_\mathrm{cen}$, Doppler broadening parameters $b_\mathrm{eff}$, and column densities $\log N$ for each absorbing component of each ion. Tables~\ref{tab:fits1} and~\ref{tab:fits2} list the parameters of the best-fit absorption profiles for all absorption components in the CGM of galaxies 1 and 2, respectively. The COS-Halos fits listed in these tables were reported in the digital form of Table~4 in \citet{Werk2013}, the limits on \ion{N}{v} were reported in Table~3 of \citet{Werk2013}, and the \ion{H}{i} fits were reported in Table~5 of \citet{Tumlinson2013}. We consider absorption lines to be ``aligned" if the velocity centroids of each line are within $\pm1\sigma$, using the formal errors returned by VPFit, of each other. We consider lines to be ``imperfectly aligned" if the velocity centroids are within $\pm3\sigma$ of each other. Other studies using COS find a systematic wavelength calibration error of $\sim7$ km s$^{-1}$ across the spectrum \citep{Tumlinson2013,Wakker2015} that may impact the velocity alignment of absorption lines. We caution that the values of $v_\mathrm{cen}$ relative to the galaxy systemic velocity may be affected by this wavelength calibration error, however, we find no evidence that wavelength calibration errors are affecting the relative velocities between absorption lines, as transitions of the same ion, or different ions with similar ionization energies that may be tracing the same gas, are well-aligned even when significantly separated in wavelength (for example, the \ion{Si}{ii} transitions of galaxy 1 at $1462-1547$\AA\ are well-aligned with each other and with the \ion{C}{ii} transition at $1538$, and in galaxy 2, the \ion{N}{ii} and \ion{O}{i} absorption lines are well-aligned despite being separated by $\sim300$\AA). We caution that while fits to saturated lines may return small formal errors, fitting Voigt profiles to saturated lines is a highly uncertain process so the errors on derived parameters from the fit may be significantly larger than the formal errors reported by VPFit. For obviously saturated lines, we report conservative lower limits on the column density by converting the measured EW to a column density using the linear part of the curve of growth.

The errors on the fitted velocity centroids are small, at the level of $1-3$ km s$^{-1}$ for clearly separated absorption components, increasing to $30-40$ km s$^{-1}$ for systems with significant overlap between the absorbing components, such as galaxy 2's \ion{O}{vi} absorption. The errors on the Doppler parameter are somewhat larger, $2-4$ km s$^{-1}$ for most components, and increasing to $10-20$ km s$^{-1}$ for galaxy 2's \ion{O}{vi} system. With the exception of galaxy 2's \ion{O}{i} absorption (0.15 dex error on one component) and \ion{O}{vi} absorption (errors of 0.29 and 0.65 dex on the two ambiguous intermediate velocity components), and of course the saturated absorption in both galaxies' CGM, the errors on the log column density for each component are only $0.02-0.07$ dex.

\subsection{Galaxy 1 at $z=0.22784$}
\label{sec:gal1_fits}

\begin{figure}
\centering
\includegraphics[width=\linewidth]{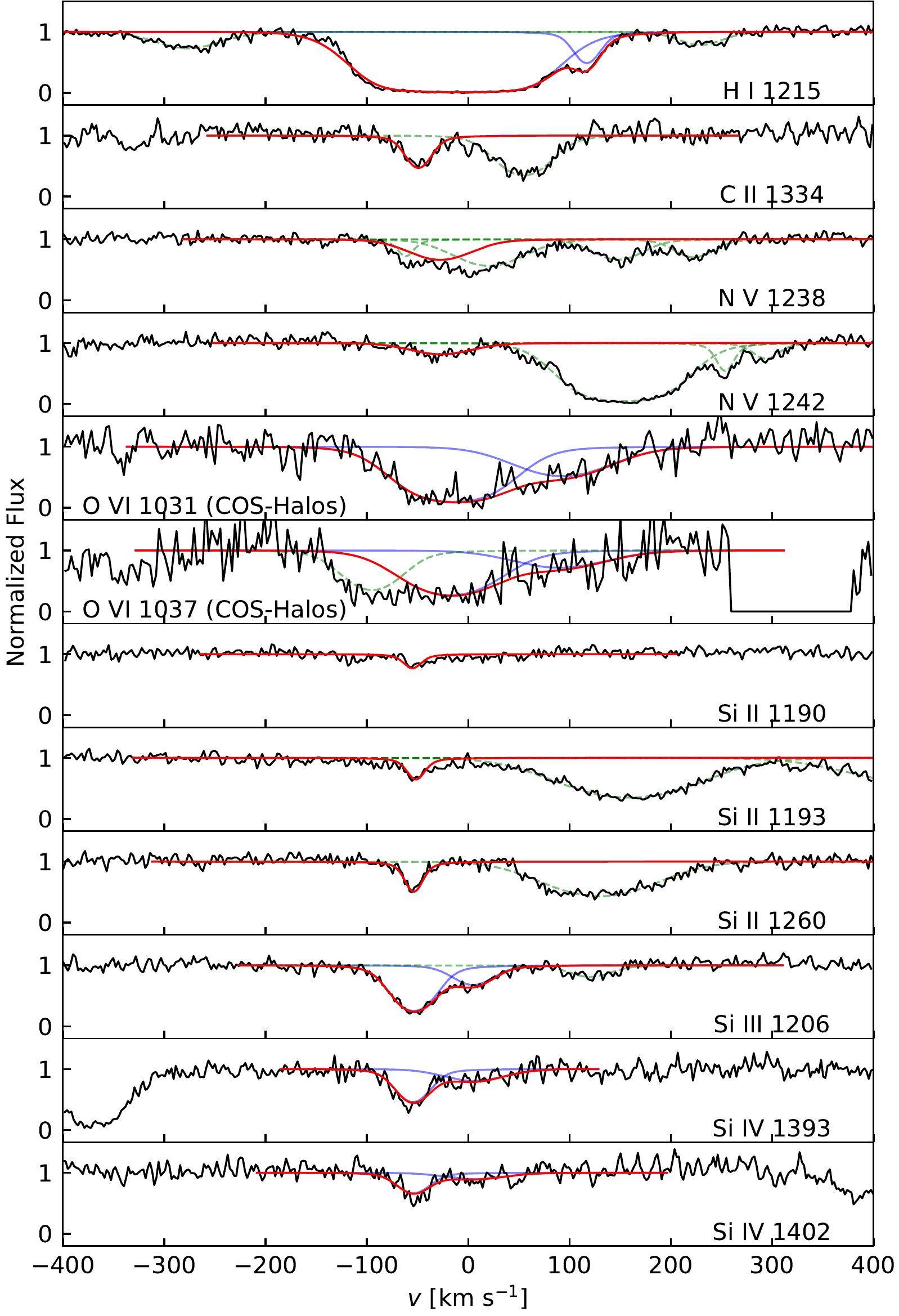}
\caption{The normalized spectrum centered on absorption lines in the CGM of the first galaxy at $z=0.22784$ as a function of velocity relative to the galaxy systemic velocity (black). Each panel shows a different absorption line, as labeled. The best-fit Voigt profiles for each absorption line are overplotted in red, and the individual absorption components are plotted in blue. For \ion{O}{vi} $\lambda1031,\lambda1037$ \AA, we show the COS-Halos spectrum and the fit to absorption in this spectrum, because our new spectrum that is the focus of this work does not cover the \ion{O}{vi} doublet at the redshift of galaxy 1.}
\label{fig:gal1_spec}
\end{figure}

Figure~\ref{fig:gal1_spec} shows all absorption lines we identify in the CGM of galaxy 1 and their best fits, as a function of line of sight velocity. A $v=0$ km s$^{-1}$ indicates that the absorption is at the systemic velocity of the galaxy, and negative velocities indicate blueshift. The spectrum is shown in black, the best-fit of the multicomponent absorption of the transition of interest is shown in red, the individual components of the best-fit absorption are shown in blue, and the best-fit to absorption components that are likely intervening absorption are shown in green dashed curves. Table~\ref{tab:fits1} reports the best-fit parameters for each ion in the CGM of galaxy 1, and Table~\ref{tab:total_columns_gal1} lists the sum of the column densities of all absorption components for each ion.

The \ion{Si}{ii} transitions show absorption at $-57$ km s$^{-1}$. \ion{Si}{iii} shows imperfectly aligned absorption at $-54$ km s$^{-1}$ and \ion{Si}{iv} shows perfectly aligned absorption at $-57$ km s$^{-1}$. \ion{Si}{iii} $\lambda 1206$ \AA\ and \ion{Si}{iv} $\lambda 1393,\lambda1402$ \AA\ also show some weak absorption perfectly aligned at $3-4$ km s$^{-1}$. \ion{C}{ii} $\lambda1334$ \AA\ shows an absorption component at $-52$ km s$^{-1}$, perfectly aligned with \ion{Si}{iii} absorption and imperfectly aligned with \ion{Si}{ii} and \ion{Si}{iv}. The absorption detected in many lines of similar ionization energies, \ion{Si}{ii}, \ion{C}{ii}, and \ion{Si}{iii} at $\sim-55$ km s$^{-1}$ may (but not necessarily) indicate $T\sim10^4$ K gas in photoionization equilibrium in the CGM of this galaxy, as is commonly assumed where such ions are detected in the CGM \citep[e.g.,][]{Prochaska2004,Tumlinson2013,Werk2014}, but we defer photoionization modeling to a future paper. There is an absorption component redshifted from \ion{Si}{iii}, but due to its large velocity ($\sim120$ km s$^{-1}$) and absence in \ion{Si}{iv} and \ion{Si}{ii} $\lambda1190$ \AA\ (we cannot tell if it is present in other transitions of \ion{Si}{ii} due to intervening absorption at that velocity), we assume this component is Lyman-$\alpha$ forest absorption and report it in Table~\ref{tab:unidentified_lines}.

We assume the \ion{N}{v} absorption is blended with other absorption lines. Because \ion{N}{v} is usually found to be a weak absorption line \citep[e.g.,][]{Werk2016}, we assume that the shallow structure just blueward of $v=0$ km s$^{-1}$ is \ion{N}{v} $\lambda1242$ \AA. This absorption is consistent with the absorption near the $\lambda1238$ \AA\ transition at this redshift, but there is significant additional absorption from intervening lines near the $\lambda1238$ \AA\ line. In fitting the doublet, we include one additional bluer component and three additional redder components surrounding the $\lambda1238$ \AA\ line, and three additional redder components near the $\lambda1242$ \AA\ line. We report the lines near the $\lambda1238$ \AA\ transition as unidentified absorption, as they are much stronger than the oscillation strength of the nearby \ion{Fe}{ii} lines would reasonably suggest (see bottom panel of Figure~\ref{fig:spectrum_chunks2}) and there are no other identified lines in this part of the spectrum. We fit them simultaneously with the doublet and report them in Table~\ref{tab:unidentified_lines}. We assume the intervening absorption near the \ion{N}{v} $\lambda1242$ \AA\ transition are \ion{Ni}{ii} from the DLA at $z=0.1140$ and \ion{Si}{ii} from the MW, and fit them simultaneously with the \ion{N}{v} doublet and report them in Table~\ref{tab:identified_lines}. Our fits produce \ion{N}{v} absorption that is weak and centered at $-31$ km s$^{-1}$. The \ion{N}{v} $\lambda1242$ \AA\ absorption has an EW of $0.07$ \AA\ and the $3\sigma$ EW limit of the spectrum at this wavelength is $0.0084$ \AA, making this a nearly $24\sigma$ detection. The \ion{N}{v} absorption imperfectly aligns with the \ion{O}{vi} absorption in the COS-Halos spectrum at $-23$ km s$^{-1}$. While the alignment of velocity centroids is not perfect, it is clear that all velocities at which there is \ion{N}{v} absorption also show \ion{O}{vi} absorption (see Figure~\ref{fig:NV_OVI_gal1}). These ions are the highest ionization states we detect in the CGM of this galaxy. We see that there are some velocities at which there is some amount of both low- and high-ion absorption (i.e. the flux is not 1), but there is no perfect or imperfect alignment between the velocity centroids of high-ions and low-ions by our definition. \emph{This indicates that the high-ions and low-ions trace kinematically separate components of the CGM.}

\ion{H}{i} is fully saturated in absorption from $-100$ km s$^{-1}$ to $75$ km s$^{-1}$, and shows three other, weaker components at $\sim-275$ km s$^{-1}$, $114$ km s$^{-1}$, and $\sim215$ km s$^{-1}$. Due to the large velocities of the components at $\sim-275$ and $\sim215$ km s$^{-1}$ and the fact that no metal lines are detected at these velocities, we do not assume that this absorption necessarily arises in the CGM of galaxy 1. Instead, these lines may be Lyman-$\alpha$ forest absorption close to but not necessarily associated with galaxy 1. We report them as unidentified lines in Table~\ref{tab:unidentified_lines}. The weaker \ion{H}{i} component at $114$ km s$^{-1}$ imperfectly aligns with the \ion{O}{vi} absorption component at $85$ km s$^{-1}$, but we do not detect absorption at this velocity in any other ion, perhaps simply because this weak absorber is tracing low column density gas. Given the large difference in ionization energy between \ion{H}{i} and \ion{O}{vi}, this component may represent comoving multiphase gas if this component of \ion{O}{vi} absorption is tracing warm, collisionally ionized gas \citep[but see][]{Stern2018}. We do not detect any metal absorption line components outside of the velocity range of \ion{H}{i} absorption.

\begin{table*}
\begin{minipage}{175mm}
\centering
\begin{tabular}{l c c c c c c c c}
 & & & & & & COS-Halos & COS-Halos & COS-Halos \\
Ion $^\mathrm{a}$ & EW (\AA)$^\mathrm{b}$ & $v_\mathrm{cen}$ (km s$^{-1}$) $^\mathrm{c}$ & $b_\mathrm{eff}$ (km s$^{-1}$) $^\mathrm{d}$ & $\log N$ (cm$^{-2}$) $^\mathrm{e}$ & $\chi^2$ $^\mathrm{f}$ & $v_\mathrm{cen}$ (km s$^{-1}$) $^\mathrm{g}$ & $b$ (km s$^{-1}$) $^\mathrm{h}$ & $\log N$ (cm$^{-2}$) $^\mathrm{i}$ \\
\hline
\ion{H}{i} 1215$^\dagger$ & 1.34 &  &  &  & $0.60$ & $-53\pm16$ & $31\pm8$ & $14.90\pm0.28$ \\
 & & $\sim-16.0$ & $\sim64.9$ & $>14.30$ &  & $18\pm14$ & $37\pm9$ & $15.25\pm0.18$ \\ 
 & & $114.3\pm0.9$ & $11.8\pm1.5$ & $13.36\pm0.03$ &  & $119\pm4$ & $18\pm6$ & $13.47\pm0.12$ \\ 
 & & & & & & $235\pm3$ & $16\pm6$ & $13.28\pm0.09$ \\
\ion{C}{ii} 1334 & 0.11 & $-52.1\pm0.9$ & $12.1\pm1.6$ & $13.83\pm0.04$ & $0.33$ & $-45.5\pm2.3$ & $12.6\pm4.1$ & $14.02\pm0.10$\\ 
\ion{N}{v} 1238 & 0.14 & $-30.5\pm2.9$ & $36.9\pm3.7$ & $13.80\pm0.04$ & $0.56$ &  &  & $<14.45$ \\ 
 & $\leq 0.0084$$^\ddag$ & $\sim84.9$ & $\sim55.7$ & $\leq 12.51$ & & & & \\
\ion{N}{v} 1242 & 0.07 &  '' & '' & '' & '' &  &  &  \\ 
\ion{O}{vi} 1031* & 0.73 & $-22.8\pm4.1$ & $51.0\pm4.2$ & $14.91\pm0.04$ & $0.42$ & $-38.0\pm6.3$ & $55.4\pm5.8$ & $14.94\pm0.06$\\ 
 & & $84.9\pm13.2$ & $55.7\pm13.5$ & $14.32\pm0.12$ &  & $74.6\pm24.7$ & $69.3\pm22.2$ & $14.44\pm0.19$\\ 
\ion{O}{vi} 1037* & 0.50 &  '' & '' & '' & '' &  &  &  \\ 
\ion{Si}{ii} 1190 & 0.03 & $-56.7\pm0.6$ & $5.2\pm0.7$ & $12.98\pm0.06$ & $0.62$ &  &  &  \\ 
\ion{Si}{ii} 1193 & 0.05 &  '' & '' & '' & '' &  &  &  \\ 
\ion{Si}{ii} 1260 & 0.07 &  '' & '' & '' & '' &  &  &  \\ 
\ion{Si}{iii} 1206 & 0.31 & $-54.2\pm1.0$ & $21.5\pm1.4$ & $13.22\pm0.02$ & $0.58$ & $-45.0\pm2.6$ & $20.0\pm4.2$ & $13.20\pm0.08$\\ 
 & & $4.5\pm2.7$ & $22.4\pm3.9$ & $12.61\pm0.05$ &  & $9.9\pm3.4$ & $8.8\pm6.8$ & $12.61\pm0.18$\\ 
\ion{Si}{iv} 1393 & 0.22 & $-57.0\pm1.1$ & $17.9\pm1.4$ & $13.28\pm0.03$ & $0.30$ &  &  &  \\ 
 & & $2.9\pm5.5$ & $36.4\pm8.1$ & $12.93\pm0.07$ &  &  &  &  \\ 
\ion{Si}{iv} 1402 & 0.13 &  '' & '' & '' & '' &  &  &  \\  
\end{tabular}
\caption{The best-fit values for each component of each ion detected in absorption in the CGM of galaxy 1, as determined by VPFit. $^\mathrm{a}$Name of the ion and transition. $^\mathrm{b}$Equivalent width of the multi-component best-fit absorption. $^\mathrm{c}$Velocity centroid of absorption relative to galaxy systemic velocity. $^\mathrm{d}$Doppler broadening width of the absorption component. $^\mathrm{e}$Log of the column density of the absorbing component. $^\mathrm{f}$The reduced $\chi^2$ of the multicomponent best fit for each ion, listed in the same row as the first component. $^\mathrm{g}$Velocity centroid of this absorption component determined by COS-Halos. $^\mathrm{h}$Doppler broadening width of this component determined by COS-Halos. $^\mathrm{i}$Log of the column density of this component determined by COS-Halos. Blank entries indicate no measurement was made, while " symbols indicate the same fitting parameters for multiple transitions of the same ion. *Note that our fits of the \ion{O}{vi} transitions were done on the COS-Halos spectrum; we performed these fits to maintain consistency with the rest of our fits and report here how our new fits in the COS-Halos spectrum compare to the originally-reported fits in the same spectrum. All COS-Halos metal line fits were originally reported in the digital form of Table~4 in \citet{Werk2013}, \ion{N}{v} upper limits were reported in Table~3 of the same paper, and \ion{H}{i} fits were reported in Table~5 of \citet{Tumlinson2013}. $^\dagger$These transitions are saturated and thus the standard Voigt fitting procedure produces highly uncertain results. We report the results of the Voigt fitting but caution the errors are likely larger than estimated by VPFit and the reported column densities are lower limits. $^\ddag$This component is a $3\sigma$ upper limit on \ion{N}{v} at the velocity of the \ion{O}{vi} component where we do not detect aligned \ion{N}{v} absorption (see \S\ref{sec:NV_OVI}).}
\label{tab:fits1}
\end{minipage}
\end{table*}

\begin{table}
\centering
\begin{tabular}{l c c}
 & & COS-Halos \\
Ion $^\mathrm{a}$ & $\log N_\mathrm{tot}$ (cm$^{-2}$) $^\mathrm{b}$ & $\log N_\mathrm{tot}$ (cm$^{-2}$) $^\mathrm{c}$ \\
\hline
\ion{H}{i} 1215 & $>14.39$ & $15.42\pm0.21$ \\ 
\ion{C}{ii} 1334 & $13.83\pm0.04$ & $14.02\pm0.10$\\ 
\ion{N}{v} 1238 & $13.80\pm0.04$ & $<14.45$ \\ 
\ion{N}{v} 1242 & " & \\ 
\ion{O}{vi} 1031 & $15.01\pm0.04$ & $15.06\pm0.06$\\ 
\ion{O}{vi} 1037 & " & \\ 
\ion{Si}{ii} 1190 & $12.98\pm0.06$ & \\ 
\ion{Si}{ii} 1193 & " & \\ 
\ion{Si}{ii} 1260 & " & \\ 
\ion{Si}{iii} 1206 & $13.32\pm0.02$ & $13.30\pm0.07$\\ 
\ion{Si}{iv} 1393 & $13.44\pm0.03$ & \\ 
\ion{Si}{iv} 1402 & " & 
\end{tabular}
\caption{The sum of the column densities of each component of each ion's absorption for galaxy 1. $^\mathrm{a}$Name of the ion and transition. $^\mathrm{b}$Sum of the column densities of each component of the best-fit absorption, as determined by VPFit. $^\mathrm{c}$Sum of the column densities of each component of the absorption reported by COS-Halos. Blank entries indicate no measurement was made, while " symbols indicate the same fit parameters for different transitions of the same ion.}
\label{tab:total_columns_gal1}
\end{table}

\subsection{Galaxy 2 at $z=0.35569$}
\label{sec:gal2_fits}

\begin{figure}
\centering
\includegraphics[width=\linewidth]{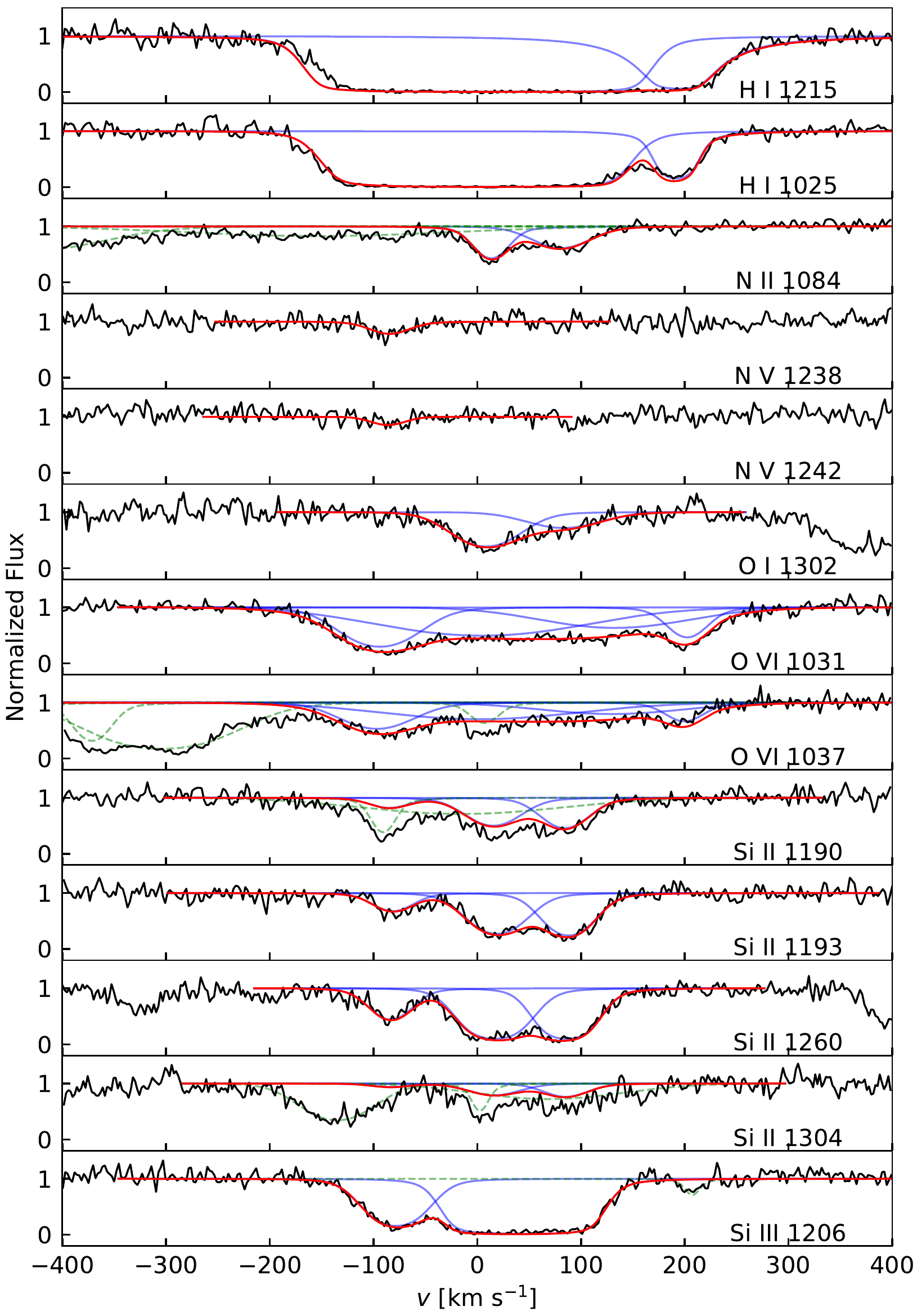}
\caption{The normalized spectrum centered on absorption lines in the CGM of the second galaxy at $z=0.35569$ as a function of velocity relative to the galaxy systemic velocity (black). Each panel shows a different absorption line, as labeled. The best-fit Voigt profiles for each absorption system are overplotted in red, and the individual absorption components are plotted in blue.}
\label{fig:gal2_spec}
\end{figure}

Figure~\ref{fig:gal2_spec} shows the absorption lines in the CGM of galaxy 2, as a function of velocity relative to the galaxy's systemic velocity, similarly to Figure~\ref{fig:gal1_spec}. Table~\ref{tab:fits2} reports the parameters of the best fits to each ion, similarly to Table~\ref{tab:fits1}. Table~\ref{tab:total_columns_gal2} lists the sum of the column densities of each absorption component for each ion, similarly to Table~\ref{tab:total_columns_gal1}.

The four \ion{Si}{ii} transitions show a multicomponent structure, with absorption features at $-86$ km s$^{-1}$, $14$ km s$^{-1}$, and $83$ km s$^{-1}$. There is significant contamination of the $\lambda1190,\lambda1304$ \AA\ transitions of this ion, so we include two additional intervening absorption components in the $\lambda1190$ \AA\ fit and three additional intervening absorption components in the $\lambda1304$ \AA\ fit. This intervening absorption near the $\lambda1190$ \AA\ transition could be due to a \ion{S}{iii} line in galaxy 2's CGM (see bottom panel of Figure~\ref{fig:spectrum_chunks3}), but that transition has a small oscillator strength ($\approx0.02$) so instead we report this intervening absorption, as well as that near the $\lambda1304$ \AA\ transition, as unidentified lines in Table~\ref{tab:unidentified_lines}. The \ion{Si}{iii} absorption is heavily saturated, but we do detect two separate, saturated components at $-78$ and $\sim44$ km s$^{-1}$. When considering the formal errors returned by VPFit, the \ion{Si}{iii} component at $-78$ km s$^{-1}$ does not quite align, even imperfectly, with the \ion{Si}{ii} absorption component at $-86$ km s$^{-1}$, but these formal errors are likely underestimated due to the saturation of this \ion{Si}{iii} component. Therefore, this \ion{Si}{iii} component and the \ion{Si}{ii} component at $-86$ km s$^{-1}$ may actually align in velocity, perfectly or imperfectly, but our definition relies on the formal errors being accurate so we are not able to make this determination in this case. In addition, because the other saturated \ion{Si}{iii} absorption component occupies the same velocity space as the other two prominent \ion{Si}{ii} absorption components, it is possible that the saturated absorption is formed by two strong absorption components that align in velocity with the two \ion{Si}{ii} absorption components, but it is impossible to distinguish this from the saturated line. We fit \ion{Si}{iii} with two components instead of three to allow for ease of comparison with the previous COS-Halos data. \ion{Si}{iii} shows an additional weak absorption component at $\sim200$ km s$^{-1}$, but the lack of this absorption component in the \ion{Si}{ii} transitions leads us to believe it is intervening absorption. There are no identified lines near this absorption component (see top panel of Figure~\ref{fig:spectrum_chunks4}), so we report it in Table~\ref{tab:unidentified_lines} as unidentified absorption. \ion{N}{ii} is also contaminated with what appears to be a very broad and shallow absorption component just blueward of the transition (see top panel of Figure~\ref{fig:spectrum_chunks2}), which we fit and report in Table~\ref{tab:unidentified_lines}. Both components of the \ion{N}{ii} and \ion{O}{i} absorption align perfectly with each other, and align imperfectly with two of the three \ion{Si}{ii} absorption components. The most-negative-velocity absorption component of \ion{Si}{ii} is not present in \ion{N}{ii} or \ion{O}{i} absorption above our EW limit.

\ion{O}{vi} is strongly absorbed and spread out in velocity space, with absorption from $-150$ to $250$ km s$^{-1}$. The number of components to fit is not immediately determinable by-eye, so we choose four, to allow for ease of comparison to the previous COS-Halos results. The \ion{O}{vi} $\lambda1037$ \AA\ transition is contaminated, so we fit an additional 3 intervening absorption components to this transition, and assume the $\lambda1031$ \AA\ transition does not contain any intervening absorption. There are a large number of metal lines from various systems along this line of sight near the \ion{O}{vi} $\lambda1037$ \AA\ transition, and we do not attempt to disentangle what line is responsible for the absorption, and report the equivalent width of the full system (minus the \ion{O}{vi} absorption) in Table~\ref{tab:identified_lines}. It is unclear how many absorption components make up the full range of \ion{O}{vi} absorption, but there are clearly separate absorption components at $-102$ km s$^{-1}$ and $195$ km s$^{-1}$. The two intermediate-velocity \ion{O}{vi} absorption components, at $-4$ and $125$ km s$^{-1}$, have very large errors on the velocity centroids and therefore are technically perfectly aligned with any other absorption in this CGM according to our definition, which we take to mean that our definition breaks down in this limit. However, it is interesting to note that there is \ion{O}{vi} absorption at any velocity where there is absorption of any other ion in this particular CGM. We detect a single, weak \ion{N}{v} absorption line at $-87$ km s$^{-1}$, with EWs of $0.06$ \AA\ at $\lambda1238$ \AA\ and $0.03$ \AA\ at $\lambda1242$ \AA\, where the $3\sigma$ EW limit is $0.021$ \AA\ at each wavelength, making these detections significant at $8\sigma$ and $4\sigma$, respectively. The absorption imperfectly aligns with the \ion{O}{vi} component at $-102$ km s$^{-1}$ and the $-79$ km s$^{-1}$ component of \ion{Si}{iii}, and perfectly aligns with the $-86$ km s$^{-1}$ component of \ion{Si}{ii}, but does not align with any absorption in \ion{O}{i} or \ion{N}{ii}. While the \ion{O}{vi} absorption covers the full range of velocities seen in absorption in the lower ionization state lines, the different relative strengths of the absorption components, and the \ion{O}{vi} component at $195$ km s$^{-1}$ that is not observed in any of the other low-ions other than \ion{H}{i}, indicates that the overall kinematic structure of this high-ionization state line is drastically different than that of the low-ionization state lines. \emph{Like galaxy 1, some of the high- and low-ion absorption components trace different gas phases that track separate components of the CGM.}

Like for galaxy 1, both the Lyman-$\alpha$ and Lyman-$\beta$ transitions of \ion{H}{i} are saturated and spread in velocity space from $\sim-150$ to $\sim250$ km s$^{-1}$, which encompasses the full range of absorption velocities of the metal ions. We detect two components in \ion{H}{i} $\lambda1025$ \AA\ and find the component at $\sim191$ km s$^{-1}$ likely aligns with the \ion{O}{vi} absorption component at $195$ km s$^{-1}$ (we cannot use our definition of perfect or imperfect alignment with the greatly underestimated errors on this component's best-fit values from VPFit), but is not seen in any of the other absorption lines of the low-ions. This component may be high-temperature gas mostly devoid of low-ions and only visible as \ion{H}{i} absorption due to the abundance of hydrogen (see Section~\ref{sec:ionization}). We defer more detailed ionization modeling to a future paper.

\begin{table*}
\begin{minipage}{175mm}
\centering
\begin{tabular}{l c c c c c c c c}
 & & & & & & COS-Halos & COS-Halos & COS-Halos \\
Ion & EW (\AA) & $v_\mathrm{cen}$ (km s$^{-1}$) & $b_\mathrm{eff}$ (km s$^{-1}$) & $\log N$ (cm$^{-2}$) & $\chi^2$ & $v_\mathrm{cen}$ (km s$^{-1}$) & $b$ (km s$^{-1}$) & $\log N$ (cm$^{-2}$) \\
\hline
\ion{H}{i} 1215$^\dagger$ & 2.39 & $\sim-1.6$ & $\sim56.1$ & $>14.55$ & $0.45$ & $-67\pm7$ & $34\pm3$ & $16.24\pm0.08$ \\ 
 & & & & & & $47\pm5$ & $33\pm1$ & $18.40\pm0.22$ \\
 & & $\sim191.1$ & $\sim7.2$ & $>14.09$ &  & $206\pm3$ & $18\pm1$ & $15.17\pm0.07$ \\ 
\ion{H}{i} 1025$^\dagger$ & 1.64 &  '' & '' & '' & '' &  &  &  \\ 
\ion{N}{ii} 1084 & 0.27 & $9.6\pm0.9$ & $16.4\pm1.5$ & $14.12\pm0.03$ & $0.59$ & $20.9\pm4.0$ & $27.6\pm6.4$ & $14.33\pm0.07$\\ 
 & & $77.0\pm1.7$ & $31.9\pm2.8$ & $14.04\pm0.03$ &  & $94.4\pm5.7$ & $18.0\pm9.7$ & $13.92\pm0.14$\\ 
 & & & & & & $182.8\pm18.7$ & $37.1\pm30.5$ & $13.62\pm0.23$\\ 
\ion{N}{v} 1238 & 0.06 & $-87.1\pm3.0$ & $22.3\pm4.4$ & $13.39\pm0.06$ & $0.34$ &  &  & $<14.00$ \\ 
 & $\leq0.021^\ddag$ & $\sim-3.4$ & $\sim121.1$ & $\leq12.84$ & & & & \\
 & $\leq0.021^\ddag$ & $\sim124.6$ & $\sim89.1$ & $\leq12.84$ & & & & \\ 
 & $\leq0.021^\ddag$ & $\sim194.8$ & $\sim19.7$ & $\leq12.84$ & & & & \\ 
\ion{N}{v} 1242 & 0.03 &  '' & '' & '' & '' &  &  &  \\ 
\ion{O}{i} 1302 & 0.44 & $6.4\pm4.3$ & $37.0\pm4.2$ & $14.65\pm0.06$ & $0.41$ & $0.2\pm4.9$ & $25.1\pm6.5$ & $14.59\pm0.10$\\ 
 & & $80.9\pm13.2$ & $43.6\pm13.0$ & $14.24\pm0.15$ &  & $77.2\pm17.5$ & $42.5\pm25.0$ & $14.22\pm0.21$\\ 
\ion{O}{vi} 1031 & 1.13 & $-101.6\pm1.9$ & $40.6\pm3.5$ & $14.48\pm0.06$ & $0.53$ & $-95.5\pm4.6$ & $41.5\pm5.5$ & $14.61\pm0.05$\\ 
 & & $-3.4_{-79.3}^{+79.2}$ & $121.1\pm39.9$ & $14.64\pm0.29$ &  & $25.5\pm9.2$ & $60.5\pm22.4$ & $14.52\pm0.12$\\ 
 & & $124.6\pm50.0$ & $89.1\pm27.6$ & $14.33\pm0.65$ &  & $117.2\pm6.9$ & $29.8\pm10.4$ & $14.21\pm0.16$\\ 
 & & $194.8\pm1.3$ & $19.7\pm2.5$ & $14.06\pm0.05$ &  & $200.7\pm2.8$ & $25.0\pm4.2$ & $14.34\pm0.05$\\ 
\ion{O}{vi} 1037 & 0.70 &  '' & '' & '' & '' &  &  &  \\ 
\ion{Si}{ii} 1190 & 0.41 & $-86.2\pm1.1$ & $22.1\pm1.7$ & $13.05\pm0.02$ & $0.38$ &  &  &  \\ 
 & & $13.8\pm1.3$ & $30.0\pm1.5$ & $13.70\pm0.02$ &  &  &  &  \\ 
 & & $82.9\pm1.1$ & $25.0\pm1.2$ & $13.70\pm0.02$ &  &  &  &  \\ 
\ion{Si}{ii} 1193 & 0.64 &  '' & '' & '' & '' &  &  &  \\ 
\ion{Si}{ii} 1260 & 0.95 &  '' & '' & '' & '' & $-97.7\pm11.8$ & $28.6\pm17.8$ & $12.99\pm0.20$\\ 
 & &  & & & & $12.4\pm12.3$ & $28.0\pm10.5$ & $13.84\pm0.17$\\ 
 & &  & & & & $76.8\pm10.0$ & $27.5\pm8.9$ & $13.91\pm0.17$\\ 
\ion{Si}{ii} 1304 & 0.18 &  '' & '' & '' & '' &  &  &  \\ 
\ion{Si}{iii} 1206$^\dagger$ & 1.24 & $-78.5\pm1.3$ & $29.2\pm1.6$ & $13.43\pm0.02$ & $0.27$ & $-77.9\pm7.1$ & $33.2\pm10.3$ & $13.58\pm0.15$\\ 
 & & $\sim43.9$ & $\sim34.5$ & $>13.62$ &  & $49.3\pm5.6$ & $27.1\pm4.9$ & $16.34\pm0.81$\\ 
 & & & & & & $215.3\pm10.1$ & $20.7\pm17.6$ & $12.62\pm0.25$
\end{tabular}
\caption{Same as Table~\ref{tab:fits1}, but for the absorption systems in the CGM of galaxy 2. All COS-Halos metal line fits were originally reported in the digital form of Table~4 in \citet{Werk2013}, \ion{N}{v} upper limits were originally reported in the digital form of Table~3 of the same paper, and \ion{H}{i} fits were reported in Table~5 of \citet{Tumlinson2013}. $^\dagger$ These transitions are saturated and thus the standard Voigt fitting procedure produces highly uncertain results. We report the results of the Voigt fitting but caution the errors are likely larger than estimated by VPFit and the reported column densities are lower limits. $^\ddag$This component is a $3\sigma$ upper limit on \ion{N}{v} at the velocity of the \ion{O}{vi} component where we do not detect aligned \ion{N}{v} absorption (see \S\ref{sec:NV_OVI}).}
\label{tab:fits2}
\end{minipage}
\end{table*}

\begin{table}
\centering
\begin{tabular}{l c c}
 & & COS-Halos \\
Ion & $\log N_\mathrm{tot}$ (cm$^{-2}$) & $\log N_\mathrm{tot}$ (cm$^{-2}$) \\
\hline
\ion{H}{i} 1215 & $>14.68$ & $18.40\pm0.22$\\ 
\ion{H}{i} 1025 & " & \\ 
\ion{N}{ii} 1084 & $14.38\pm0.02$ & $14.53\pm0.06$\\ 
\ion{N}{v} 1238 & $13.39\pm0.06$ & $<14.00$\\ 
\ion{N}{v} 1242 & " & \\ 
\ion{O}{i} 1302 & $14.79\pm0.06$ & $14.74\pm0.09$\\ 
\ion{O}{vi} 1031 & $15.03\pm0.18$ & $15.05\pm0.05$\\ 
\ion{O}{vi} 1037 & " & \\ 
\ion{Si}{ii} 1190 & $14.05\pm0.01$ & \\ 
\ion{Si}{ii} 1193 & " & \\ 
\ion{Si}{ii} 1260 & " & $14.20\pm0.11$\\ 
\ion{Si}{ii} 1304 & " & \\ 
\ion{Si}{iii} 1206 & $>13.84$ & $16.34\pm0.81$\\ 
\end{tabular}
\caption{Same as Table~\ref{tab:total_columns_gal1}, but for the second galaxy's absorption.}
\label{tab:total_columns_gal2}
\end{table}

\subsection{Comparison to Previous Fits}

In Tables~\ref{tab:fits1}-\ref{tab:total_columns_gal2}, the results from fitting the absorption lines as determined by the COS-Halos study are included, where available. We focus on line detections in the COS-Halos survey, but include the non-detection upper limits on \ion{N}{v} because we discuss the column density ratio $\log (N_\mathrm{NV}/N_\mathrm{OVI})$ extensively below. We find similar velocity centroids as found by COS-Halos, within the wavelength calibration error of COS of $\sim10$ km s$^{-1}$ \citep{Tumlinson2013}. The values of the Doppler parameter $b_\mathrm{eff}$ we find are similar to those found in the COS-Halos study, with the largest differences between our measurements and the COS-Halos measurements occurring for the saturated or otherwise ambiguous absorption components with the largest errors.

Many of the column densities we recover are similar to those reported by COS-Halos, both the column densities of individual absorption components and the summed column density across all components. Our lower limit on the column density for \ion{Si}{iii} in galaxy 2's CGM is $1.07$ dex smaller than the COS-Halos measurement, but we expect that the saturation of this line makes measuring its column density highly uncertain from formal Voigt profile fitting. Our lower limit is, however, consistent with their measurement.

Despite some differences in the velocity, width, and strength of the multiple components of \ion{O}{vi} absorption in galaxy 2's CGM between our fits and COS-Halos, the summed column densities across all components for both studies match within the errors. COS-Halos reports a detection of a third \ion{N}{ii} absorption component with relatively small $N_\mathrm{NII}=13.62\pm0.23$ cm$^{-2}$ at $v=182.8\pm18.7$ km s$^{-1}$ in galaxy 2's CGM \citep[see digital form of Table~4 in][]{Werk2013}, which was perhaps due to a spurious fluctuation in the low SNR COS-Halos spectrum. With our higher SNR spectrum, it is clear that there is no significant \ion{N}{ii} absorption with EW $>0.015$ \AA\ at this velocity (see Figure~\ref{fig:gal2_spec}).

In general, the errors on our derived absorption parameters are smaller than those reported by COS-Halos, due to our higher SNR spectrum. For galaxy 1, which has fewer saturated or multicomponent lines, our errors on $v_\mathrm{cen}$ are $\sim1/2$ those of COS-Halos, the errors on $b$ are $\sim1/2-2/3$ those of COS-Halos, and the errors on $\log N$ are $0.06-0.13$ dex smaller than COS-Halos. Galaxy 2 has more saturated or extended multicomponent absorption where the separation of the components is not clear, such as the \ion{O}{vi} absorption, where our errors on the fitted parameters are similar to those reported by COS-Halos. Our fits perform much better than COS-Halos for galaxy 2's \ion{Si}{ii}, as we are able to fit multiple transitions of the ion simultaneously. In these cases, our errors on $v_\mathrm{cen}$ and $b$ are a factor of $5-10$ smaller than COS-Halos. Our errors on the column densities are $0.15-0.5$ dex smaller than COS-Halos reports for \ion{Si}{ii}. Increased precision of the fits on non-saturated absorbers may prove to be helpful for constraining models and is an added benefit of high SNR spectra, along with the most important benefit of detecting weak absorption in the CGM discussed in \S\ref{sec:new_lines}.

\section{Discussion}
\label{sec:discussion}

\subsection{Ionization Processes}
\label{sec:ionization}

\subsubsection{Galaxy 1 at $z=0.22784$}

There is absorption in \ion{C}{ii}, \ion{Si}{ii}, \ion{Si}{iii}, and \ion{Si}{iv} in the CGM of galaxy 1 that is perfectly or imperfectly aligned (see Figure~\ref{fig:gal1_spec}), indicating these ions may arise in comoving, and possibly cospatial, gas. Low-ionization gas is typically assumed to be photoionized at a low temperature of $T\sim10^4$ K, \citep[e.g.,][]{Prochaska2004} and \ion{Si}{iv} may arise due to either photoionization or collisional ionization. The alignment of \ion{Si}{iv} absorption components with lower ion absorption components may indicate that \ion{Si}{iv} is photoionized, rather than collisionally ionized. Without detailed photoionization modeling, which will be performed in a forthcoming paper, we cannot speculate on the density or ionization parameter.

The most-positive velocity component of the \ion{O}{vi} absorption at $85$ km s$^{-1}$ may be aligned with the \ion{H}{i} absorption at $114$ km s$^{-1}$, but we cannot use our definition of alignment in this case due to the likely underestimated errors returned by VPFit for saturated lines. If these two absorption components are aligned, and if they trace the same gas at the same temperature, then the velocity width of the \ion{H}{i} component should be consistent with the temperature inferred from the \ion{O}{vi} absorption component. If the \ion{O}{vi}-absorbing gas were in CIE, its temperature would be expected to be $\sim10^{5.5}$ K \citep{Gnat2007,Oppenheimer2013}, which would lead to a thermal broadening of \ion{H}{i} of $b_\mathrm{eff}\approx 70$ km s$^{-1}$. The \ion{H}{i} absorption component at $114$ km s$^{-1}$ has a width of $b_\mathrm{eff}=12$ km s$^{-1}$, so if the \ion{H}{i} and \ion{O}{vi} absorption components trace the same phase of gas, it cannot be in CIE at the expected temperature.

\citet{Stern2018} propose a model in which \ion{O}{vi} absorption traces low-temperature photoionized gas beyond the accretion shock in a galaxy's halo. They estimate the expected column density ratios of $N_\mathrm{NV}/N_\mathrm{OVI}$, $N_\mathrm{SiIII}/N_\mathrm{OVI}$, and $N_\mathrm{HI}/N_\mathrm{OVI}$ in photoionized, low-density gas. To compare with this model, we assume the \ion{O}{vi} and \ion{N}{v} absorption components at $-31$ and $-23$ km s$^{-1}$, respectively, are tracing the same gas as the \ion{C}{ii}, \ion{Si}{ii}, \ion{Si}{iii}, and \ion{Si}{iv} components near $\sim-55$ km s$^{-1}$, even though these lines are not aligned, either perfectly or imperfectly, with our definition. For these components, we find $N_\mathrm{NV}/N_\mathrm{OVI}=0.08$ and $N_\mathrm{SiIII}/N_\mathrm{OVI}=0.02$. $N_\mathrm{HI}/N_\mathrm{OVI}$ is difficult to calculate for only a single component of the \ion{O}{vi} absorption because the \ion{H}{i} absorption is saturated. However, the range of velocities extended by the \ion{O}{vi} absorption is similar to that of the \ion{H}{i} absorption component at $\sim-16$ km s$^{-1}$, so for a very rough estimate we assume our lower limit on the \ion{H}{i} column density for this component is its actual value, and that this \ion{H}{i} component is tracing the same gas as the components of the other ions considered here. This gives $N_\mathrm{HI}/N_\mathrm{OVI}=0.25$. The $N_\mathrm{NV}/N_\mathrm{OVI}$ value is consistent with the expected column density ratio for solar metallicity or somewhat sub-solar metallicity gas in the model of \citet{Stern2018} (see their Figure~7), but the $N_\mathrm{SiIII}/N_\mathrm{OVI}$ value requires significantly super-solar metallicity to match this model. If $N_\mathrm{HI}$ is roughly 1.5 dex larger than our lower limit, a solar metallicity gas in photoionization equilibrium could give rise to our measured $N_\mathrm{SiIII}/N_\mathrm{OVI}$ value, but then the $N_\mathrm{NV}/N_\mathrm{OVI}$ value would imply a sub-solar metallicity to match the \citet{Stern2018} model and the assumption that these ions trace the same gas would be invalidated. This implies that if the absorption we detect arises from photoionized gas outside the accretion shock, it must be quite metal-enriched, which may be unlikely for gas at such large distances from galaxies. In addition, the fact that the absorption components we explore here do not actually align in velocity indicates that the \ion{O}{vi} and \ion{N}{v} absorption is likely a different kinematic component of galaxy 1's CGM than the \ion{Si}{ii}, \ion{C}{ii}, \ion{Si}{iii}, and \ion{Si}{iv} components.

\subsubsection{Galaxy 2 at $z=0.35569$}

The strong lines of \ion{Si}{iv} fall outside the wavelength range of our spectrum for galaxy 2, but we see that absorption in \ion{Si}{ii} and \ion{Si}{iii} lines are aligned, although the likely underestimated errors on the saturated \ion{Si}{iii} components make determining the quality of alignment impossible. \ion{N}{ii} and \ion{O}{i} are perfectly aligned with each other and imperfectly aligned with two of the absorption components of \ion{Si}{ii}, and possibly \ion{Si}{iii}. This may indicate that each of these lines are tracing the same comoving and possibly cospatial gas, similar to galaxy 1. The presence of multiple low-ionization state lines in galaxy 2's CGM that may trace the same gas indicates the gas may be photoionized, as we saw in galaxy 1 above. Again, detailed photoionization modeling is necessary to derive densities and ionization parameters, which will be presented in a forthcoming paper.

We do not detect multiple absorption components in \ion{H}{i} $\lambda1215$\ to compare to the absorption components in other ions, but there is one component in \ion{H}{i} $\lambda1025$\ at $191$ km s$^{-1}$ that may align with the \ion{O}{vi} absorption at $\sim195$ km s$^{-1}$, but again the quality of the alignment is difficult to guess due to the saturation of \ion{H}{i}. If the comoving gas traced by both this \ion{H}{i} component and this \ion{O}{vi} component is cospatial, in CIE, and at a constant temperature, then again we would expect the thermal broadening of the \ion{H}{i} component to be $b_\mathrm{eff}\approx 70$ km s$^{-1}$, consistent with the temperature of $\sim10^{5.5}$ K. Similarly to galaxy 1, the \ion{H}{i} component that aligns with the \ion{O}{vi} component in galaxy 2's CGM has a Doppler broadening width too small to be produced by high-temperature gas: $b_\mathrm{eff}\sim8$ km s$^{-1}$. We conclude that if the gas producing the absorption components in \ion{H}{i} and \ion{O}{vi} at $\sim190$ km s$^{-1}$ are cospatial, this gas cannot be in CIE.

We can again compare to the model of \citet{Stern2018}. For galaxy 2, the \ion{N}{v} absorption is imperfectly aligned with the \ion{O}{vi} component at $-102$ km s$^{-1}$ and the \ion{Si}{iii} component at $-79$ km s$^{-1}$. If we assume these absorption components are tracing the same gas, then we calculate $N_\mathrm{NV}/N_\mathrm{OVI}=0.08$ and $N_\mathrm{SiIII}/N_\mathrm{OVI}=0.09$. To get a very rough estimate of $N_\mathrm{HI}/N_\mathrm{OVI}$, we assume the column density of the saturated \ion{H}{i} absorption component at $\sim-2$ km s$^{-1}$ is equal to the lower limit we find, and then split it evenly among the three lowest-velocity components of \ion{O}{vi} absorption, because this \ion{H}{i} component and these three \ion{O}{vi} components show absorption over a similar range of velocities. We caution that this is an extremely rough estimate. We find that $N_\mathrm{HI}/N_\mathrm{OVI}=0.39$. As we found for galaxy 1, the $N_\mathrm{SiIII}/N_\mathrm{OVI}$ ratio implies a significantly super-solar metallicity to match the model of \citet{Stern2018}, while the $N_\mathrm{NV}/N_\mathrm{OVI}$ ratio implies solar metallicity, suggesting that these ions are not tracing the same gas. Again, if the \ion{H}{i} column density is 1.5 dex higher than our lower limit, the $N_\mathrm{SiIII}/N_\mathrm{OVI}$ ratio aligns with the solar metallicity model of \citet{Stern2018}, but then the $N_\mathrm{NV}/N_\mathrm{OVI}$ ratio implies a significantly sub-solar metallicity. Like galaxy 1, the fact that the rest of the \ion{O}{vi} absorption shows a different kinematic structure than the other absorption lines indicates that perhaps some of the \ion{O}{vi} components could be tracing a different kinematic component of galaxy 2's CGM than some of the low-ion lines, even if one of the \ion{O}{vi} components is tracing gas outside an accretion shock.

\subsection{N V and O VI}
\label{sec:NV_OVI}

\citet{Indebetouw2004} present a list of various non-equilibrium ionization processes collected from the literature and the $\log (N_\mathrm{NV}/N_\mathrm{OVI})$ ratios expected from each process. Here, we calculate the \ion{N}{v} to \ion{O}{vi} column density ratios for our two galaxies' CGM and compare to the ratios in \citet{Indebetouw2004} to estimate the ionization process for these high ions. We also compare to \citet{Wakker2012} and \citet{Bordoloi2017} for more recent determinations of the high-ionization state metal column densities expected from a flow of radiatively cooling gas, to \citet{Werk2016} and \citet{McQuinn2018} for other possible non-equilibrium processes, and to other measured values of the $\log (N_\mathrm{NV}/N_\mathrm{OVI})$ ratio in the Milky Way halo \citep{Wakker2012}, other low-$z$ galaxies \citep{Werk2013}, low-$z$ intergalactic medium absorbers \citep{Burchett2015}, high-$z$ galaxies \citep{Lehner2014}, and high-$z$ damped Ly-$\alpha$ absorbers \citep{Fox2007}. While we generally focus on detections rather than upper limits, we do include the upper limits on $\log (N_\mathrm{NV}/N_\mathrm{OVI})$ measured by COS-Halos for the two galaxies discussed in this paper, for a direct comparison with previous results.

\subsubsection{Galaxy 1 at $z=0.22784$}

\begin{figure}
\centering
\includegraphics[width=\linewidth]{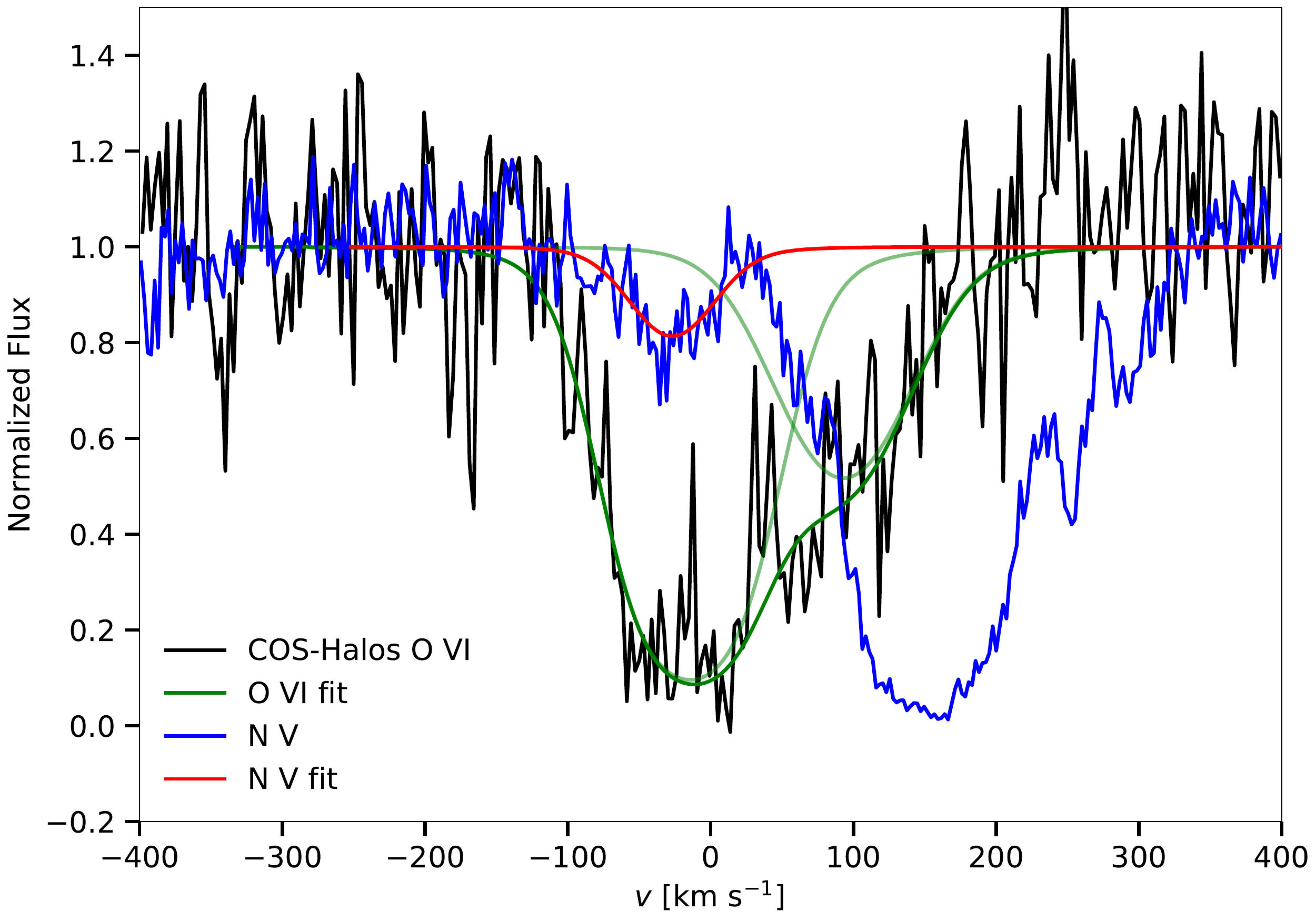}
\caption{Galaxy 1's \ion{N}{v} $\lambda1242$ \AA\ absorption (blue) and best fit from our high SNR spectrum (red) overplotted in velocity space on the COS-Halos \ion{O}{vi} $\lambda1031$ \AA\ absorption (black) and our best two-component fit to the COS-Halos spectrum (dark green for full profile, light green for components). While the best-fit velocity centroids of these absorption components are imperfectly aligned, the velocity range of the \ion{N}{v} absorption is fully covered by the velocity range of the strongest \ion{O}{vi} absorption component. The strong absorption redward of the \ion{N}{v} line is likely a combination of \ion{Ni}{ii} $\lambda13270$ \AA\ absorption in the DLA at $z=0.1140$ and \ion{Si}{ii} $\lambda1526$ \AA\ absorption in the MW halo at $z=0$.}
\label{fig:NV_OVI_gal1}
\end{figure}

In Figure~\ref{fig:NV_OVI_gal1}, we overplot the \ion{N}{v} absorption from our spectrum at $\lambda1242$ \AA\ and the \ion{O}{vi} absorption from the COS-Halos spectrum at $\lambda1031$ \AA\ to show that while the best-fit velocity centroids of these absorption components are imperfectly aligned, the velocity range of the \ion{N}{v} absorption is fully covered by the strongest \ion{O}{vi} absorption component. We calculate the ratio of \ion{N}{v} to \ion{O}{vi} column densities for each absorption component in \ion{O}{vi}, using an upper limit on \ion{N}{v} detection for those \ion{O}{vi} components that are not paired with \ion{N}{v} absorption at similar systemic velocities. We use the linear part of the curve of growth to convert our $3\sigma$ EW limits into $3\sigma$ column density upper limits and find a limiting column density for \ion{N}{v} non-detection of $\log (N_\mathrm{NV})\leq12.51$ for galaxy 1. While there is an intervening absorber in the spectrum near \ion{N}{v} $\lambda1242$ \AA\ at the velocity that would align with the second \ion{O}{vi} absorption component, the absorption near \ion{N}{v} $\lambda1238$ \AA\ is not consistent with absorption at this velocity so we consider this a non-detection and use the detection limit to place a limit on $\log (N_\mathrm{NV}/N_\mathrm{OVI})$.

The log ratio of the column densities for the detected \ion{N}{v} absorption and imperfectly-aligned \ion{O}{vi} is $\log (N_\mathrm{NV}/N_\mathrm{OVI})=-1.11\pm0.06$. The $3\sigma$ upper limit on the log-ratio of the column densities for the \ion{N}{v} non-detection at the velocity of the other \ion{O}{vi} component is $\log (N_\mathrm{NV}/N_\mathrm{OVI})\lesssim-1.81$. Figure~\ref{fig:ionratio} reproduces Figure~1 from \citet{Indebetouw2004}, with the vertical red line with shading noting the detected value of $\log (N_\mathrm{NV}/N_\mathrm{OVI})$ and its error for galaxy 1. The non-detection is plotted only in the bottom panel as an open red circle with a left-pointing arrow. The detection line passes through both the conductive heating \citep[of spherical clouds in a hot media;][]{Boehringer1987} and the radiative cooling regimes \citep[of Galactic fountain gas;][]{Shapiro1991}, crosses and triangles, respectively, as reported by \citet{Indebetouw2004}. The non-detection value does not correlate with any regime collected by \citet{Indebetouw2004}, but does align with some upper limit detections from other systems (see \S\ref{lab:meas_comp}). The difference in the values of the \ion{N}{v} to \ion{O}{vi} ratio between the detected and non-detected components may indicate that there are multiple ionization processes for \ion{N}{v} and \ion{O}{vi} within the same halo, but there is a large amount of intervening absorption near both lines of the \ion{N}{v} doublet that may be hiding a second absorption component that aligns with the second \ion{O}{vi} absorption component. To distinguish between the non-equilibrium collisional ionization processes of radiative cooling and conductive heating would require further ion coverage, preferably of \ion{C}{iv}, but \citet{Werk2016} present the argument that turbulent mixing \citet[in the model of][]{Gnat2010} produces too-small \ion{O}{vi} column densities, so we focus on radiatively cooling flows as our preferred non-equilibrium ionization mechanism of the gas traced by \ion{N}{v} and \ion{O}{vi}.

\citet{Wakker2012} find that a radiatively cooling flow model can produce a ratio $\log (N_\mathrm{NV}/N_\mathrm{OVI})\sim-1.1$ and individual column densities $\log N_\mathrm{NV}\approx13.8$ and $\log N_\mathrm{OVI}\approx14.9$ if the flow is isochoric (i.e. if the flow is not constrained by external pressure from a hot halo) and has a velocity of $\approx160$ km s$^{-1}$, assuming the sightline is parallel to the direction of the flow, (see their Figure~15). Thus, the \ion{N}{v}-\ion{O}{vi} column density ratio for galaxy 1 could be produced by a cooling flow of intermediate speed, even though the ratio is somewhat higher than suggested for a radiatively cooling flow from \citet{Indebetouw2004} in Figure~\ref{fig:ionratio}. The radiatively cooling flow reported by \citet{Indebetouw2004} was calculated by \citet{Shapiro1991} for a Galactic fountain with flow velocity $100$ km s$^{-1}$. \citet{Bordoloi2017} also find a radiatively cooling flow can be responsible for our measured values of $\log N_\mathrm{NV}\approx13.8$ and $\log N_\mathrm{OVI}\approx14.9$, but the velocity required for the flow is higher, $\approx200$ km s$^{-1}$. Their model also constrains the temperature of the cooling gas to $\sim10^{5.3-5.5}$ K (see their Figure~5).

\subsubsection{Galaxy 2 at $z=0.35569$}

\begin{figure}
\centering
\includegraphics[width=\linewidth]{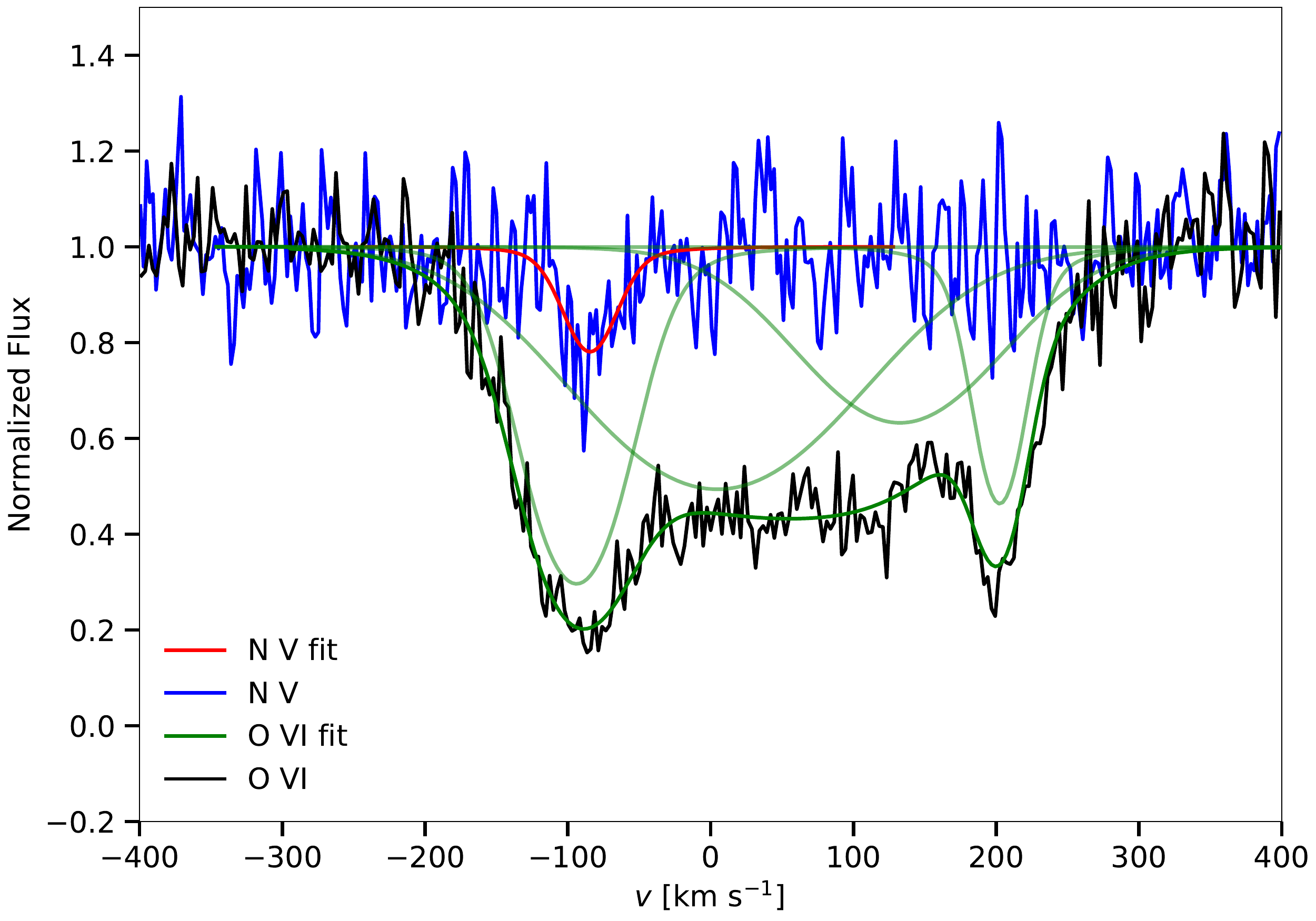}
\caption{Galaxy 2's \ion{N}{v} $\lambda1238$ \AA\ absorption (blue) and best fit (red) overplotted in velocity space on the \ion{O}{vi} $\lambda1031$ \AA\ absorption (black) and best fit (dark green for full profile, light green for components), both from our high SNR spectrum. The strongest \ion{O}{vi} absorption component is imperfectly aligned with the \ion{N}{v} absorption.}
\label{fig:NV_OVI_gal2}
\end{figure}

Our spectrum covers both the \ion{N}{v} and \ion{O}{vi} transitions at the redshift of galaxy 2. In Figure~\ref{fig:NV_OVI_gal2}, we overplot the \ion{N}{v} $\lambda1238$ \AA\ (the stronger of the \ion{N}{v} doublet lines) and \ion{O}{vi} $\lambda1031$ \AA\ absorption (the only \ion{O}{vi} line that does not have intervening absorption) in velocity space from our spectrum, along with the best multi-component fits. While \ion{O}{vi} has extended, multi-component absorption at a wide range of velocities, we detect weak \ion{N}{v} absorption in only a single component at $v\approx-87$ that imperfectly aligns with the $-102$ km s$^{-1}$ component of the \ion{O}{vi} absorption, which is the strongest component. We calculate $\log (N_\mathrm{NV}/N_\mathrm{OVI})$ for the \ion{N}{v} detection and the imperfectly aligned \ion{O}{vi} component, and find a $3\sigma$ upper limit on $\log N_\mathrm{NV}\leq12.84$ for the \ion{N}{v} non-detections at $\lambda1238$ \AA.

For the one \ion{O}{vi} component with imperfectly aligned \ion{N}{v} detected absorption, we find $\log (N_\mathrm{NV}/N_\mathrm{OVI})=-1.09\pm0.08$. For the other three \ion{O}{vi} components at $-3$, $125$, and $195$ km s$^{-1}$, we place $3\sigma$ upper limits of $\log (N_\mathrm{NV}/N_\mathrm{OVI})\lesssim-1.8$, $\lesssim-1.49$, and $\lesssim-1.22$, respectively. In Figure~\ref{fig:ionratio}, we plot the one detected \ion{N}{v} component's ratio as a vertical orange dashed line with shading to indicate the error on this ratio, and the three limits are plotted only in the bottom panel as open orange circles with left-pointing arrows. The one detection of this ratio for galaxy 2 is strongly consistent with the one detection value of this ratio for galaxy 1. Without a measurement of \ion{C}{iv} absorption, the models reported by \citet{Indebetouw2004} cannot be directly compared to our data to predict which ionization process produces the \ion{N}{v} and \ion{O}{vi} gas, but the detection and one of the limits on the \ion{N}{v} to \ion{O}{vi} column density ratio for galaxy 2 are consistent with either radiative cooling or conductive heating, as for galaxy 1. We again favor radiative cooling over conductive heating for the same reason as for galaxy 1, that conductive heating predicts far lower \ion{O}{vi} column densities than we measure. The spread in the \ion{N}{v} to \ion{O}{vi} ratio between the different components is much larger than the spread in this ratio from a single type of non-equilibrium ionization. For example, the extent of the conductive heating points (crosses) from \citet{Indebetouw2004} is $\sim0.4$ dex in the \ion{N}{v} to \ion{O}{vi} ratio, while the spread between all components of galaxy 2's absorption is $\gtrsim0.7$ dex. It is clear that the variety of values for the \ion{N}{v} to \ion{O}{vi} ratio cannot be produced by a single ionization process, but instead either multiple ionization processes must be taking place in the same halo or, if these ions arise due to radiatively cooling flows, they must be flowing at different velocities.

As for galaxy 1, the isochoric radiatively cooling flow model of \citet{Wakker2012} can produce the column density ratio observed for galaxy 2. Despite the similar column density ratios between the two galaxies, galaxy 2 has lower values of the column density for the imperfectly aligned detections of \ion{N}{v} and \ion{O}{vi}. In the model of \citet{Wakker2012}, a lower velocity cooling flow is required to produce these column densities than for galaxy 1, of $\approx50$ km s$^{-1}$. The cooling flow model of \citet{Bordoloi2017} produces $\log N_\mathrm{NV}\approx13.4$ and $\log N_\mathrm{OVI}\approx14.5$ when the flow has a velocity of $\sim80$ km s$^{-1}$ and a temperature of $\sim10^{5.3-5.5}$ K (see their Figure~5). Therefore, the \ion{N}{v} and \ion{O}{vi} in galaxy 2's CGM may arise due to a radiatively cooling flow of gas, albeit at a lower velocity than galaxy 1.

\subsubsection{Comparison to Other Measurements}
\label{lab:meas_comp}

\begin{figure*}
\begin{minipage}{175mm}
\centering
\includegraphics[width=\linewidth]{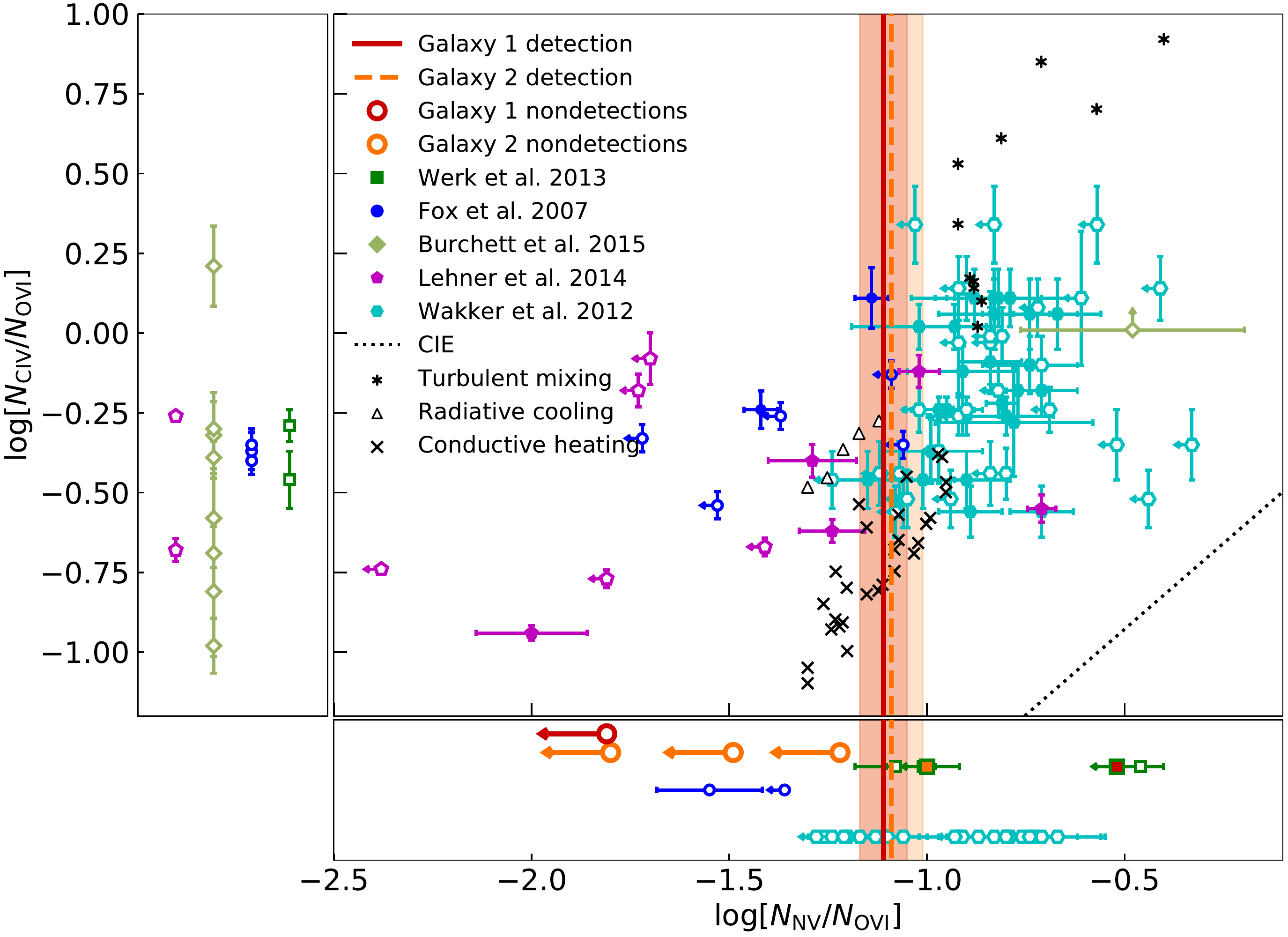}
\caption{In the large panel, the log of the ratio of \ion{C}{iv} column density to \ion{O}{vi} column density vs. the log of the ratio of \ion{N}{v} to \ion{O}{vi} column densities. Bottom and left panels are reserved for points with a measurement in only one of the two plotted column density ratios, with the exception of our own measured values, which are also plotted in the large panel. Our detected measurements of $\log (N_\mathrm{NV}/N_\mathrm{OVI})$ are plotted as vertical lines with shading indicating the error on the measurement: solid red line for galaxy 1 and dashed orange line for galaxy 2. Upper limits from non-detections are plotted as open circular points in only the bottom panel, again red for galaxy 1 and orange for galaxy 2. Black points are reproduced from Figure~1 of \citet{Indebetouw2004}: crosses show regions of column density ratios expected for conductive heating, asterisks show regions of turbulent mixing, triangles show regions of radiative cooling, and the diagonal dotted line in the bottom right shows ratios expected from CIE. Colored points indicate measured values of the column density ratios collated from various other systems: the CGM of low-$z$, $\sim L^\star$ galaxies \citep[dark green squares][]{Werk2013}, high-$z$ damped Ly-$\alpha$ absorbers \citep[blue circles][]{Fox2007}, low-$z$ intergalactic medium absorbers \citep[light green diamonds][]{Burchett2015}, the CGM of high-$z$ galaxies \citep[magenta pentagons][]{Lehner2014}, and the Milky Way halo \citep[cyan hexagons][]{Wakker2012}. Filled colored points indicate a measurement for both column density ratios, whereas an open colored point indicates a measurement for one column density ratio and a limit for the other. The COS-Halos upper limits on $\log (N_\mathrm{NV}/N_\mathrm{OVI})$ are mostly neglected to avoid over-crowding the plot with points that have neither \ion{N}{v} nor \ion{C}{iv} detections, but we do plot the COS-Halos upper limits for the two galaxies studied here as the larger green squares with red or orange face colors for galaxy 1 or galaxy 2, respectively, in the bottom plot.}
\label{fig:ionratio}
\end{minipage}
\end{figure*}

In addition to showing our measured values of $\log (N_\mathrm{NV}/N_\mathrm{OVI})$ and the models collected by \citet{Indebetouw2004}, Figure~\ref{fig:ionratio} also shows a collection of $\log (N_\mathrm{NV}/N_\mathrm{OVI})$ and $\log (N_\mathrm{CIV}/N_\mathrm{OVI})$ measurements from other studies as the colored points: the CGM of low-$z$, $\sim L^\star$ galaxies \citep[dark green squares][]{Werk2013}, high-$z$ damped Ly-$\alpha$ absorbers \citep[blue circles][]{Fox2007}, low-$z$ intergalactic medium absorbers \citep[light green diamonds][]{Burchett2015}, the CGM of high-$z$ galaxies \citep[magenta pentagons][]{Lehner2014}, and the Milky Way halo \citep[cyan hexagons][]{Wakker2012}. The upper limits on $\log(N_\mathrm{NV}/N_\mathrm{OVI})$ measured by COS-Halos for these two galaxies are included for comparison in Figure~\ref{fig:ionratio} as larger green squares with red and orange face colors indicating galaxy 1 and galaxy 2, respectively, in the bottom panel.

The COS-Halos \ion{N}{v} non-detection for galaxy 1 produced an upper limit of $\log (N_\mathrm{NV}/N_\mathrm{OVI})<-0.52$, which could have been consistent with gas in CIE, so our detection shows a remarkable improvement over the previous best limit. For galaxy 2, the COS-Halos upper limit is $\log (N_\mathrm{NV}/N_\mathrm{OVI})<-1.0$, so our detection does not show as strong of an improvement over the upper limit as for galaxy 1. Unless the column density of \ion{C}{iv} that is not covered by our spectrum is an order of magnitude lower than the lowest values found in the CGM of other galaxies plotted in Figure~\ref{fig:ionratio}, CIE appears to be ruled out as the ionization mechanism for the gas traced by \ion{N}{v} and \ion{O}{vi} for both galaxies\footnote{Note that \citet{Werk2016} concluded that high-temperature CIE could produce the $N_\mathrm{NV}/N_\mathrm{OVI}$ upper limits they measured, and because our detections are consistent with their upper limits, our detected $N_\mathrm{NV}/N_\mathrm{OVI}$ could also be produced by high-temperature gas in CIE. However, this would produce a very low value of $N_\mathrm{CIV}$ \citep{Gnat2007}, significantly different than other, similar galaxies' CGM, so we do not favor this ionization mechanism even though our $N_\mathrm{NV}/N_\mathrm{OVI}$ is consistent with it and we do not measure \ion{C}{iv} to break the degeneracy.}. For galaxy 1, this reduces the number of possible ionization mechanisms, as the COS-Halos limit was consistent with CIE. Unlike galaxy 1, our detected $N_\mathrm{NV}/N_\mathrm{OVI}$ for galaxy 2 is not far below the upper limit reported by COS-Halos, indicating that it is impossible to predict by how much upper limits are over-estimating the true values of $N_\mathrm{NV}$. While additional ion coverage, particularly of \ion{C}{iv}, would be immensely helpful to determining the ionization process of the gas, even without \ion{C}{iv}, a \ion{N}{v} detection goes further toward classifying the ionization process of the gas than an upper limit does. The values of our $N_\mathrm{NV}$ detections are similar to many of the $N_\mathrm{NV}$ upper limits in other galaxies' CGM from \citet{Werk2016}, but more \ion{N}{v} detections are needed to ensure that the upper limits are not drastically over-estimating $N_\mathrm{NV}$ (like we found for galaxy 1) and to confirm the ionization mechanism of gas traced by \ion{N}{v}.

Our measured values for $\log (N_\mathrm{NV}/N_\mathrm{OVI})$ (solid red and dashed orange vertical lines with shaded errors in Figure~\ref{fig:ionratio}) most closely align with the measurements of the CGM of low-$z$, $\sim L^\star$ galaxies from \citet{Werk2013} (dark green squares), some of the measurements of the CGM of high-$z$ galaxies from \citet{Lehner2014} (magenta pentagons), some of the measurements of high-$z$ damped Ly-$\alpha$ absorbers from \citet{Fox2007} (blue circles), and the low-$\log (N_\mathrm{NV}/N_\mathrm{OVI})$ edge of the Milky Way halo measurements from \citet{Wakker2012} (cyan hexagons). This is consistent with the findings of \citet{Bordoloi2017} that many of these measurements can be explained by radiatively cooling gas flows, just as we find our measurements can. In the case of those measurements of the CGM near galaxies, like our measurements, the flowing gas may be fueled by galactic winds. High-velocity cooling gas $\gtrsim500$ km s$^{-1}$ may be hot, fast galactic winds radiatively cooling as they adiabatically expand out of the galaxy \citep{Thompson2016}. \citet{Lochhaas2018} found that fast galactic winds slow down to $\lesssim100-200$ km s$^{-1}$ and can radiatively cool after shocking on the CGM gas, consistent with observations of low-velocity cooling gas.

Our preferred gas process that gives rise to \ion{N}{v} and \ion{O}{vi} absorption, namely radiatively cooling flows, is consistent with the findings of \citet{Werk2016,McQuinn2018} \citep[but see also][for a picture in which \ion{O}{vi} arises from photoionized gas]{Stern2016,Stern2018}, who test various gas processes that could give rise to \ion{N}{v} and \ion{O}{vi} absorption and conclude that the most likely scenario is a very massive, radiatively cooling flow. This ``flow'' could arise due to galactic winds from the galaxy, or it could be a warm gas phase that traces the virial temperature of $L^\star$ galaxies ``sloshing'' and continuously cooling and being re-heated by energetic feedback within the halo. This latter scenario may be more likely than radiatively cooling galactic winds because of the very large mass fluxes of radiatively cooling gas required. They base the majority of these conclusions on \ion{O}{vi} kinematics and column densities, as only 3 spectra in the COS-Halos survey detect \ion{N}{v}. \citet{Werk2016} use upper limits $\log (N_\mathrm{NV}/N_\mathrm{OVI})\sim-1$ to $-1.5$ to explore the ionization process of \ion{N}{v} and \ion{O}{vi} absorbing gas, but it is clear that detections, rather than upper limits, are necessary to distinguish between multiple scenarios that produce similar ionic column density ratios. For example, the spread in $\log (N_\mathrm{NV}/N_\mathrm{OVI})$ between different photoionization models is smaller than the spread in upper limits of this ratio \citep[see figure~12 of][]{Werk2016}. High SNR spectra with detections of \ion{N}{v}, as we have presented here, can help narrow down the parameter space of possible ionization processes and gas kinematics that give rise to high-ionization state absorbing gas in the CGM.

\section{Summary and Conclusions}
\label{sec:summary}

We have identified and fit absorption lines in a high SNR HST COS spectrum toward the quasar SDSS J1009+0713, which is located behind the CGM of two galaxies at $z=0.228$ and $z=0.356$. We report on the column densities, velocities, and Doppler broadening parameters of the absorption lines in the CGM of each of these galaxies, and show that our high SNR spectrum allows us to more accurately measure the absorbing gas than in previous, shallower surveys. Our main results are as follows:
\begin{enumerate}
\item We identify new absorption previously not reported in the CGM of both galaxies: weak \ion{N}{v} absorption in both galaxies' CGM, \ion{Si}{ii} absorption in galaxy 1's CGM, and clearer decomposition of \ion{O}{vi} absorption component in galaxy 2's CGM (\S\ref{sec:new_lines}).
\item We identify new absorption at $z=0$ corresponding to absorption in the MW's CGM as well as at $z=0.456$, corresponding to absorption of gas surrounding the quasar. The newly-identified quasar absorption includes a fast outflow of \ion{O}{vi} up to $\sim1100$ km s$^{-1}$ (Figure~\ref{fig:spectrum_chunks2}).
\item The different kinematic structure of low- and high-ionization lines in both galaxies' CGM indicates these ions are produced in different components of the CGM, are not comoving, and likely not cospatial (\S\ref{sec:lines}).
\item We see similar kinematic absorption structure in multiple low-ionization state lines (\ion{Si}{ii}, \ion{C}{ii}, \ion{N}{ii}, \ion{Si}{iii}); these low-ions may be photoionized (\S\ref{sec:ionization}) and will be modeled in a forthcoming paper. The narrow features of \ion{H}{i} absorption that may be aligned in velocity with components of \ion{O}{vi} absorption show that these \ion{O}{vi} and \ion{H}{i} components are not tracing the same phase of gas in CIE because \ion{H}{i} is not observed to be thermally broadened to the expected temperature.
\item The column density ratios of the detected \ion{N}{v} and \ion{O}{vi} we find in the CGM of both galaxies are consistent with those predicted by a radiatively cooling flow of gas traveling at $\sim50-150$ km s$^{-1}$, and these column density ratios are also similar to those measured in the CGM of other galaxies as well as in the Milky Way halo. Radiatively cooling gas traveling at this velocity can be produced by fast galactic winds shocking on CGM gas (\S\ref{sec:NV_OVI} and Figure~\ref{fig:ionratio}) or by ``sloshing'' of cooling gas within a halo \citep{Werk2016,McQuinn2018}.
\item The nondetections of \ion{N}{v} produce drastically different \ion{N}{v} to \ion{O}{vi} column density ratios, even within a single galaxy's CGM, which may indicate that either \ion{N}{v} and \ion{O}{vi} are produced in multiple different ionization processes within the same CGM (\S\ref{sec:NV_OVI} and Figure~\ref{fig:ionratio}) or, if the ionization process is radiatively cooling flows, there must be multiple cooling flows of drastically different velocities within each galaxy's CGM.
\item We report on unidentified lines in the spectrum and find that, if we assume they are Ly-$\alpha$ forest absorption lines, they follow the expected number density of Ly-$\alpha$ forest absorbers from both simulations of the intergalactic medium and observational studies at this redshift (Appendix \ref{sec:Lya_forest}). The number of these lines that are broad also follow the expected number of BLAs at this redshift (Appendix \ref{sec:BLA}).
\end{enumerate}

We have shown that a high SNR spectrum can improve errors on measured column densities by $\sim0.1$ dex and improve absorption line velocity centroids and Doppler broadening widths by a factor of $\sim2$. These precise measurements reduce the uncertainty on ionization process for the high-ionization state metals, such as \ion{N}{v}, for which measurements were previously dominated by upper limits of non-detections. Such deep spectra are crucial for uncovering weak absorbers that may give insight to the kinematics of a variety of CGM gas, a complementary approach to modeling the strongest CGM absorbers that trace the bulk of CGM material and can be studied with lower SNR spectra. We will report on more detailed ionization modeling of the absorption lines in this spectrum in a future paper. While broad and shallow surveys give a statistical understanding of the CGM, deep spectra like the one reported in this paper probe the potentially different processes occuring within a single galaxy's CGM. Additional deep spectra will improve our understanding of the gas phases and hydrodynamic processes of the CGM.

\section*{Acknowledgments}

CL thanks John Chisholm and Adam Leroy for fruitful discussion. The authors thank the anonymous referee for providing useful suggestions that greatly improved the quality of the paper. We gratefully acknowledge support for our HST program (number GO-14708), provided by NASA through a grant from the Space Telescope Science Institute, which is operated by the Association of Universities for Research in Astronomy, Inc., under NASA contract NAS5-26555. C. L. is supported in part by NSF Grant \#1516967. Y. K. acknowledges support from the grant PAPIIT 106518.

\appendix

\section{\lowercase{d}$N$/\lowercase{d}$z$ of Ly-$\alpha$ Absorbers}
\label{sec:Lya_forest}

Table~\ref{tab:unidentified_lines} lists all unidentified absorption systems in our spectrum. We checked whether these unidentified absorbers were Ly-$\alpha$ or Ly-$\beta$ absorption by searching for absorption at the corresponding redshift of other Lyman series lines, in both our spectrum and the previously-obtained COS-Halos G130M spectrum. We find that if the unidentified systems at $1768.4$, $1767.5$, $1727.0$, $1718.9$, $1706.9$, $1671.8$, $1638.9$, $1626.0$, and $1556.3$ \AA\ are Ly-$\alpha$ lines, then there is corresponding Ly-$\beta$ absorption at the wavelengths $1492.5$, $1491.3$, $1457.2$, $1451.4$, $1440.0$, $1409.9-14.11.1$ \AA\ in our spectrum and at the wavelengths $1383$, $1372$, and $1313$ \AA\ in the COS-Halos spectrum. Of those in our spectrum, these potential Ly-$\beta$ lines are either listed as unidentified absorbers in Table~\ref{tab:unidentified_lines} or are absorbers blended with the CGM absorption of galaxies 1 or 2 that we fit simultaneously with the CGM absorption in those galaxies. We cannot test the assumption that unidentified absorbers with wavelengths blueward of the absorber at $1466.0$ \AA\ are Ly-$\alpha$ because the rest of the Lyman series lines fall outside the range of both our spectrum and the COS-Halos spectrum. Instead, we test these lines by searching for absorption at the wavelengths where Ly-$\alpha$ or Ly-$\gamma$ would be if these were Ly-$\beta$. We find that these lines cannot be Ly-$\beta$ because there is no detectable absorption at the wavelengths expected for Ly-$\alpha$, and the EWs of these absorbers, if they were Ly-$\beta$, would predict EWs of the Ly-$\alpha$ lines above our $3\sigma$ limit. Our spectrum extends fully to the quasar redshift at $z=0.456$, so any detected Ly-$\beta$ absorption would also be detected in Ly-$\alpha$. There are three ambiguous unidentified absorbers at $1716.1$ \AA, $1641.6$ \AA, and $1636.8$ \AA\ that are weak so additional Lyman series lines, if these are Ly-$\alpha$, would be undetected.

Out of the 25 independent absorbers we detect (i.e., not double-counting those we identify as Ly-$\alpha$ and Ly-$\beta$ pairs), we only have evidence that 9 are true Ly-$\alpha$ absorbers, because the corresponding Ly-$\beta$ absorption is consistent with other absorbers in our spectrum or the COS-Halos spectrum. For another 13 absorbers, we have some evidence that they are \emph{not} Ly-$\beta$ by ruling out absorption at corresponding Lyman series lines, so we assume they are Ly-$\alpha$ lines, but this is not necessarily the case if there are additional unidentified metal lines in the spectrum. The final 3 absorbers are too weak to detect additional Lyman series lines regardless of if they are Ly-$\alpha$ or Ly-$\beta$ and so are ambiguous.

If we assume all 25 independent unidentified absorbers in the spectrum are Ly-$\alpha$ forest absorbers, we can compare the number of absorbers in our spectrum to the expected number at the spectrum's redshift, albeit with large errors due to possible misidentification of absorbers and the fact that we have only a single spectrum over which to count Ly-$\alpha$ forest absorbers. The spectrum covers Ly-$\alpha$ redshifts from 0.15 to 0.45, but some of this redshift range is ``covered up" due to absorption of other lines that are not Ly-$\alpha$ forest: the absorbers in the CGM of either galaxy 1 or 2 and the metal lines of the quasar, the MW, or the DLA at $z=0.1140$. These other lines cover $\sim20\%$ of the spectrum, so the Ly-$\alpha$ forest redshift range, $\Delta z$, probed by this spectrum is $\Delta z\sim0.24$. Restricting the column density of absorbers to $N(\mathrm{Ly}\alpha)\geq 13$, which is nearly $11\sigma$ above our average EW limit of detection, we find 29 individual absorption components in the spectrum. Restricting to $N(\mathrm{Ly}\alpha)\geq14$ reduces the number of individual absorption components to 6. We calculate d$N$/d$z$ by dividing the number of absorbers by the redshift extent of the spectrum that is not covered by absorption of other lines, $\Delta z=0.24$, and compute the error on d$N$/d$z$ using Poisson small-number statistics. We find d$N$/d$z=120_{-33}^{+45}$ for $N(\mathrm{Ly}\alpha)\geq 13$ and d$N$/d$z=25_{-14}^{+24}$ for $N(\mathrm{Ly}\alpha)\geq14$, where the errors are $\pm2\sigma$. Using the power-law distribution of d$N$/d$z$ found by \citet{Danforth2016}, we should expect $\sim6$ absorbers for $N(\mathrm{Ly}\alpha)\geq14$ and $\sim27$ absorbers for $N(\mathrm{Ly}\alpha)\geq 13$ given our $\Delta z\sim0.24$ at a redshift of $z\sim0.2-0.3$, which is consistent with the numbers we find within the errors provided by Poisson statistics. Note, however, that counting the number of Ly-$\alpha$ absorbers in a single spectrum, even if we were confident that all of these absorbers are in fact Ly-$\alpha$, is likely to stochastically differ from the numbers counted in \citet{Danforth2016}, who used a significantly larger number of sightlines. We do not claim that our single sightline, in which possibly several absorbers have been misidentified as Ly-$\alpha$, can make any predictions of d$N$/d$z$, but we do find consistency with larger and more careful Ly-$\alpha$ forest surveys.

\begin{figure}
\centering
\includegraphics[width=\linewidth]{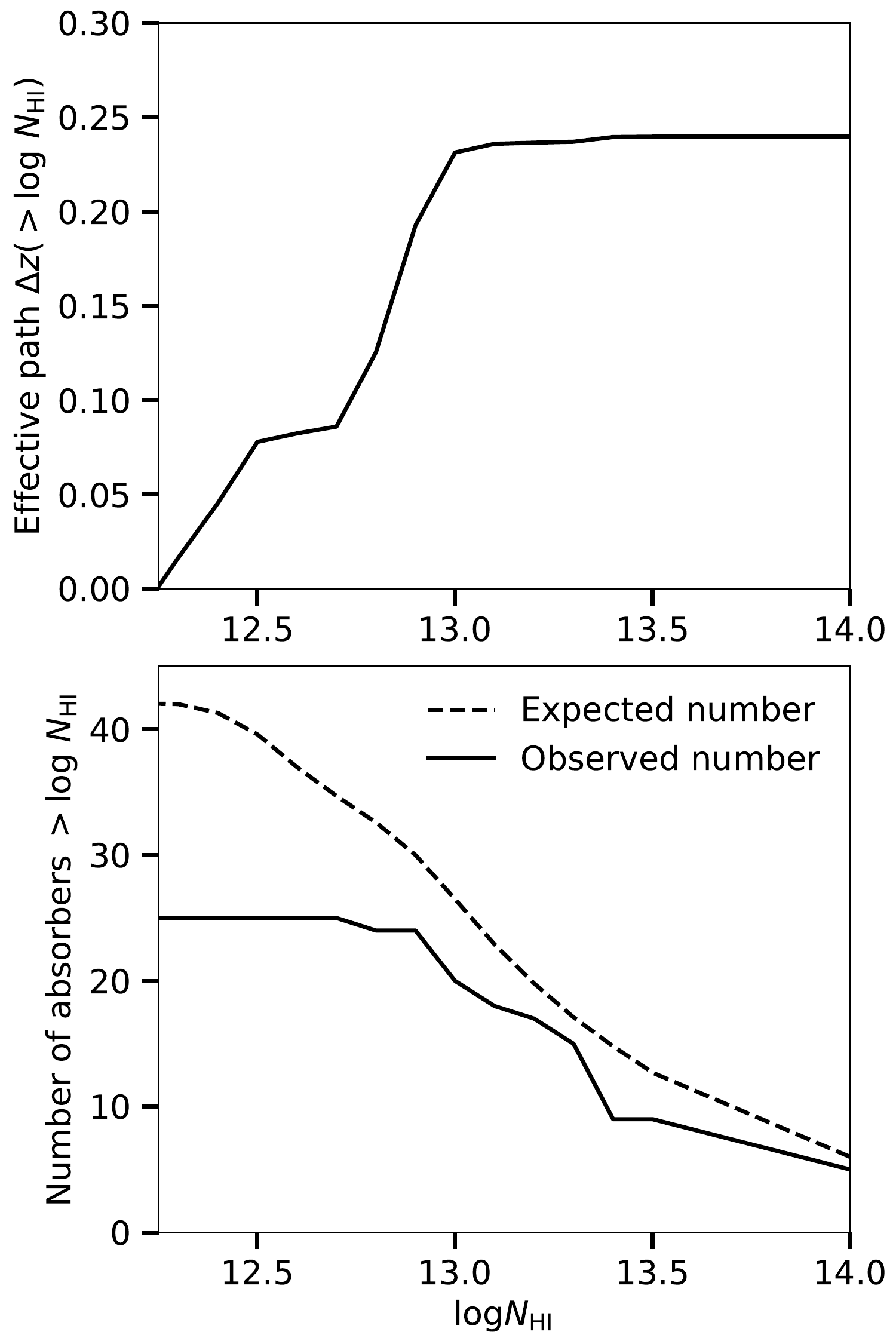}
\caption{Upper panel: Effective pathlength available for a secure $4\sigma$ detection as function of limiting column density $\log N_\mathrm{HI}$ for the unidentified absorbers assumed to be Ly$\alpha$. Note that we have excluded wavelength areas covered by already identified absorption for this estimate. Lower panel: Observed total number of absorbers as function of limiting column density (solid line), and expected total number (dashed line). The calculation of the expected numbers relies upon the effective path, and assumes a column density distribution function in the form of a power law, appropriately adjusted to our redshift regime as prescribed by \citet{Danforth2016}. We do not show errors because we have not performed a thorough accounting and analysis of low-column density absorbers in our spectrum, but the number of absorption systems we find agrees within $1\sigma$ with the expectation from \citet{Danforth2016} for $N_\mathrm{HI}>12.8$ and disagrees for $\log N_\mathrm{HI}<12.8$. The probability of detecting only one absorber within $12.3 < \log N_\mathrm{HI} < 12.8$ is $<1\%$ if the power-law distribution found by \citet{Danforth2016} continues unchanged to these low column densities (see text).}
\label{fig:number_colden}
\end{figure}

The high data quality allows us to probe the low column density range for the unidentified absorber distribution \citep[see also][]{Williger2010, Danforth2016}. The upper panel of Fig.~\ref{fig:number_colden} shows the effective pathlength available in our spectrum as function of the minimum column density $\log N_\mathrm{HI}$, when demanding a $4\sigma$ detection criterion, following \citet{Danforth2016}. As detailed above, this pathlength excludes all parts of the spectrum covered by already identified absorption. Essentially, we are complete to a limit of $\log N_\mathrm{HI}\sim 13.0$, but even down to $\log N_\mathrm{HI}\sim 12.5$ there is a non-negligible fraction of the spectrum that enables the secure detection of such weaker lines. Assuming thus a column density distribution function in the form of a power law, where both the normalisation $N(z)$ and the slope $\beta(z)$ have been redshift-evolved in the manner appropriate for our redshift coverage as prescribed by \citet{Danforth2016}, allows us to estimate the number of absorbers within given column density intervals in our spectrum. The lower panel of Fig.~\ref{fig:number_colden} shows the cumulative number of absorbers expected above a limiting column density from this calculation (dashed line). For absorbers above $\log N_\mathrm{HI} \sim 12.8$, the observed number of absorbers (solid line) agrees well with expectations within $1\sigma$ errors (not shown), however, for weaker absorbers there appears to be an observed lack: out of the 9.4 expected absorbers within $12.3 < \log N_\mathrm{HI} < 12.8$, we only find 1 (and that one appears in a blend, making the column density estimate rather uncertain). The likelihood to obtain one or zero absorbers in this case, $P(x\leq1.0, \mu = 9.4)$, is 0.086\%. If any of our absorbers have been mis-identified as Ly-$\alpha$ when they are in fact metal lines or other Lyman series lines, our results are even more discrepant with what is expected. While we caution that the main focus of this paper was neither the detection nor the analysis of such weak absorbers, and hence there could be a small number of weak absorbers that may have escaped our scrutiny, we note that this apparent dearth of absorption with $\log N_\mathrm{HI} <\sim 12.8$ could indicate a shallower slope than expected in that regime for the column density distribution function and represents an important reason for obtaining further high SNR spectra. Note, however, that \citet{Williger2010,Danforth2016} did not find any evidence for a break in the column density distribution at low column densities using data of similar SNR, so our dearth of low column absorbers may be a statistical fluctuation or due to incompleteness in our found absorbers.

\section{Broad Ly-$\alpha$ Absorbers}
\label{sec:BLA}

Broad Ly-$\alpha$ absorbers (BLAs) are regions of high column density neutral hydrogen gas with a high temperature $T\sim10^{5.5}$ K that thermally broadens the absorption line. At these temperatures, the fraction of hydrogen in the neutral state is very low and their optical depth is spread over a wide range in velocity, so BLAs are weak and require high SNR spectra to find. However, they are useful for probing the warm-hot intergalactic medium because the thermal broadening is expected to be larger than any non-thermal broadening \citep{Richter2006} so a clear measurement of the temperature can be made.

We search for potential BLAs by fitting each component of Ly-$\alpha$ absorption with a Voigt profile using VPFit and identifying any single-component absorber with a Doppler $b_\mathrm{eff}$ parameter greater than 40 km s$^{-1}$. These absorbers are marked in bold in Table~\ref{tab:unidentified_lines} and Table~\ref{tab:BLAs} lists the wavelength at which we find them, their redshift assuming they are Ly-$\alpha$ lines, their Doppler $b_\mathrm{eff}$ parameters, and their column densities.

\begin{table}
\centering
\begin{tabular}{l c c c}
 & & & $\log N_{\mathrm{Ly}\alpha}\ ^\mathrm{d}$ \\
Wavelength $^\mathrm{a}$ (\AA) & $z\ ^\mathrm{b}$ & $b_\mathrm{eff}\ ^\mathrm{c}$ (km s$^{-1}$) &  (cm$^{-2}$) \\
\hline
1466.0 & 0.20585 & 77.1 & 14.07 \\
1468.6 & 0.20821 & 167.7 & 13.63 \\
1475.5 & 0.21371 & 58.7 & 13.68 \\
1476.4 & 0.21444 & 59.8 & 13.80 \\
1477.6 & 0.21544 & 120.1 & 13.31 \\
1521.5 & 0.25129 & 40.7 & 13.56 \\
1556.3 & 0.28014 & 40.3 & 13.84 \\
1613.3 & 0.32745 & 123.7 & 13.76 \\
1629.1 & 0.34011 & 41.3 & 13.41 \\
1727.0 & 0.42057 & 56.8 & 13.70 \\
1768.4 & 0.45492 & 91.1 & 13.61
\end{tabular}
\caption{Broad Ly-$\alpha$ absorbers in our spectrum, defined as Ly-$\alpha$ absorbers with Doppler $b_\mathrm{eff}>40$ km s$^{-1}$. $^\mathrm{a}$Wavelength of the BLA. $^\mathrm{b}$Redshift of the absorption. $^\mathrm{c}$The Doppler $b_\mathrm{eff}$ broadening parameter. $^\mathrm{d}$The log of the column density.}
\label{tab:BLAs}
\end{table}

At a SNR level of $30$, \citet{Tepper-Garcia2012} predict d$N$/d$z\sim28\pm5$ for simple (single-component) BLAs, using hydrodynamical simulations of the intergalactic medium. We find 11 in our spectrum that covers $\Delta z=0.24$, giving d$N$/d$z=45_{-19}^{+25}$, where we report $\pm2\sigma$ Poisson errors. Our SNR is $\sim25$, so our finding is not quite within $1\sigma$ of the prediction; we find slightly more BLAs than we expect. However, 3 of the BLAs in our spectrum have a Doppler $b_\mathrm{eff}$ parameter within 2 km s$^{-1}$ of the BLA identifying limit of $b_\mathrm{eff}\geq 40$ km s$^{-1}$, and the typical errors on $b_\mathrm{eff}$ in our fits are $\sim2-4$ km s$^{-1}$. Some of these BLAs may be mis-identified, if their actual $b_\mathrm{eff}$ values are smaller than we find, which would reduce our calculated d$N$/d$z$ to $33_{-13}^{+22}$, consistent with the number of BLAs predicted for our SNR from hydrodynamical simulations.

The d$N$/d$z$ of BLAs in our spectrum is also somewhat larger than the numbers found in quasar spectra in observational studies. \citet{Richter2006} and \citet{Lehner2007} both find a d$N$/d$z\sim20-25$ and \citet{Danforth2010} find d$N$/d$z=18\pm11$. Our findings for our spectrum are marginally consistent within the errors with the d$N$/d$z$ found by these other studies. This may be some indication that some of the absorption in our spectrum is not Ly-$\alpha$ as we assume, or it may simply be a statistical fluctuation of our single sight line compared to the averages from larger surveys. Additional high SNR spectra would be instrumental to the search for BLAs.

None of the BLAs we identify are associated with either galaxy's CGM probed by our spectrum, so they may be tracing the low redshift intergalactic medium, which is expected to be warm-hot, or could be located in other nearby galaxies' CGM. Those with the largest column densities $\log N_\mathrm{HI}\gtrsim14$ are close to saturated, so may be misidentified as broad absorption. The \ion{H}{i} absorption we detect in the CGM of the two galaxies is fully saturated so if there are any BLA components, they are blended with the saturated \ion{H}{i} and we cannot separate them out.

\end{document}